\documentclass[11pt]{article}

\usepackage{a4}
\usepackage[mathscr]{eucal}
\usepackage{citesort}
\usepackage{url}
\usepackage{amssymb,latexsym}
\usepackage{float}

\DeclareFontFamily{OT1}{rsfs}{} 
\DeclareFontShape{OT1}{rsfs}{m}{n}{ <-7> rsfs5 <7-10> rsfs7 <10-> rsfs10}{} 
\DeclareMathAlphabet{\mycal}{OT1}{rsfs}{m}{n} 
\def\scri{{\mycal I}}%
\def\scrip{\scri^{+}}%
\def\scrp{{\mycal I}^{+}}%
\def\Scri{\scri}

\newcommand{\mK}{anti-de Sitter}
\newcommand{\Omegak}{\Omega_k}

\newcommand{\ce}{\check{e}}
\newcommand{\halpha}{{\hat \alpha}}
\newcommand{\hrho}{{\hat \rho}}
\newcommand{\hsigma}{{\hat \sigma}}

\newcommand{\zero}{{\hat 0}}
\newcommand{\one}{{\hat 1}}
\newcommand{\two}{{\hat 2}}
\newcommand{\three}{{\hat 3}}
\newcommand{\A}{{\hat A}}

\newcommand{\ha}{{\hat a}}
\newcommand{\ta}{{\tilde a}}
\newcommand{\hb}{{\hat b}}
\newcommand{\hc}{{\hat c}}
\newcommand{\hr}{{\hat r}}
\newcommand{\hs}{{\hat s}}
\newcommand{\hmu}{{\hat \mu}}
\newcommand{\hnu}{{\hat \nu}}
\newcommand{\hatc}{{\hat c}}
\newcommand{\hA}{{\hat A}}

\newcommand{\hi}{{\hat \imath}}

\newcommand{\newk}{\tilde k}
\newcommand{\fourg}{{}^4g} 

\def\greaterthansquiggle{\raise.3ex\hbox{$>$\kern-.75em\lower1ex\hbox{$\sim$}}}
\def\lessthansquiggle{\raise.3ex\hbox{$<$\kern-.75em\lower1ex\hbox{$\sim$}}}

\newcommand{\rds}{r_{\partial\Sigma}}
\newcommand{\bnabla}{\overline{\nabla}}
\newcommand{\pSinfty}{\partial_{\infty}\Sigma}
\newcommand{\bg}{\overline{g}}
\newcommand{\bart}{\overline{t}}
\newcommand{\ch}{\check{h}} 
\newcommand{\bM}{\overline{M}}
\newcommand{\fbg}{{}^4\overline{g}}
\newcommand{\fg}{{}^4{g}}
\newcommand{\fgp}{{}^4{g'}}
\newcommand{\sg}{{}^{\scri}\!\!{g}}
\newcommand{\sR}{{}^{\scri}\!{R}}
\newcommand{\bh}{\overline{h}}
\newcommand{\pSo}{\partial\Sigma} 
\newcommand{\pSm}{\partial_1\Sigma} 
\newcommand{\pSi}{\partial_i\Sigma} 
\newcommand{\be}{\begin{equation}}
\newcommand{\bS}{\overline{\Sigma}}
\newcommand{\bSo}{\overline{\Sigma}_0}
\newcommand{\gK}{generalized Kottler}

\newcommand{\GHn}{Hawking}
\newcommand{\mgh}{M_{Haw}}
\newcommand{\mghu}{\mgh(u)}
\newcommand{\mghp}{\mgh(\psi)}
\def\varOmega{{\mathit{\Omega}}}
\def\varomega{{\mathit{\omega}}}
\newcommand{\bDel}{\overline\Delta}
\newcommand{\bDelta}{\overline\Delta}
\newcommand{\bD}{\overline D}
\newcommand{\bR}{\overline R}
\newcommand{\pM}{\scri}

 \newtheorem{Theorem} {Theorem} [section]
 \newtheorem{Corollary} [Theorem] {Corollary}
 \newtheorem{Lemma} [Theorem] {Lemma}
 \newtheorem{Proposition} [Theorem] {Proposition}

 \newtheorem{Definition}[Theorem]{Definition}

\newcommand{\myremark}{\begin{remark}\rm} 
\newcommand{\myendremark}{\end{remark}}

\newcommand{\kind}{i}                         

\newcommand{\cR}{{\mathcal{R}}}
\newcommand{\mham}{M_{Ham}}

\newcommand{\cU}{\mycal U}
\newcommand{\cV}{\mycal V}

\newcommand{\cNX}{{{\mathscr  N}[X]}{}}

\newcommand{\cD}{{\mathscr  D}}

\newcommand{\cO}{{\mathscr  O}}

\newcommand{\text}[1]{\mbox{\rm #1}}
\newcommand{\qed}{\hfill $\Box$ \medskip} 
\newcommand{\proof}{\noindent {\sc Proof:\ }}

\newcommand{\bn}{\overline{n}}
\newcommand{\bp}{\overline{p}}

\newcommand{\Mtwo}{{}^2M}

\newcommand{\Sext}{\Sigma_{{\text{\scriptsize\rm ext}}}}

\newcommand{\rd}{{\rm d}}
\newcommand{\beq}{\begin{equation}}

\newcommand{\eeq}{\end{equation}}
\newcommand{\ee}{\end{equation}}
\newcommand{\beqa}{\begin{eqnarray}}
\newcommand{\eeqa}{\end{eqnarray}}
\newcommand{\beqan}{\begin{eqnarray*}}
\newcommand{\eeqan}{\end{eqnarray*}}
\newcommand{\ba}{\begin{array}}
\newcommand{\ea}{\end{array}}

\def\au{{\setbox0=\hbox{\lower1.36775ex\hbox{''}\kern-.05em}\dp0=.36775ex\hs
kip0pt\box0}}
\def\ao{{}\kern-.10em\hbox{``}}

{\catcode `\@=11 \global\let\AddToReset=\@addtoreset}
\AddToReset{equation}{section}

\newcounter{mnotecount}[section]

\newcommand{\Chi}{\Xi}

\newcommand{\eq}[1]{(\ref{#1})}

\newcommand{\gf}{f} 

\newcommand{\commentout}[1]{}

\newcommand{\bea}{\begin{eqnarray}}
\newcommand{\eea}{\end{eqnarray}}
\newcommand{\beaa}{\begin{eqnarray*}}
\newcommand{\eeaa}{\end{eqnarray*}}

\newcommand{\N}{{\mathbb N}}
\newcommand{\R}{{\mathbb R}}

\newcommand{\backg}{b}\newcommand{\bmetric}{{\backg}}
\newcommand{\hyp}{\Sigma}
\newcommand{\ourU}{\mathbb U}

\renewcommand{\include}{\input}

\begin{document}

\title{Towards the classification of static vacuum 
  spacetimes with  negative cosmological constant}

\author{Piotr T.\ Chru\'sciel \thanks{Supported in part by KBN grant
    \# 2 P03B 073 15. \emph{E--mail}: Chrusciel@Univ-Tours.fr} \\ 
  D\'epartement de
  Math\'ematiques\\ Facult\'e des Sciences\\ Parc de Grandmont\\
  F37200 Tours, France\\ \\
  Walter Simon\thanks{Supported by Jubil\"aumsfonds der
    \"Osterreichischen Nationalbank, project \# 6265, and by a grant
    from R\'egion Centre, France.  \emph{E--mail}:
    simon@ap.univie.ac.at}
  \\Institut f\"ur theoretische Physik \\
  Universit\"at Wien,\\ Boltzmanngasse 5,\\ A-1090 Wien, Austria}

 \maketitle 
\begin{abstract}
  {We present a systematic study of static solutions of the vacuum
    Einstein equations with negative cosmological constant which
    asymptotically approach the generalized Kottler
    (``Schwarzschild---anti-de Sitter'') solution, within (mainly) a
    conformal framework.  We show connectedness of conformal infinity
    for appropriately regular such space-times. We give an explicit
    expression for the Hamiltonian mass of the (not necessarily
    static) metrics within the class considered; in the static case we
    show that they have a finite and well defined \GHn\ mass. We prove
    inequalities relating the mass and the horizon area of the
    (static) metrics considered to those of appropriate reference \gK\ 
    metrics. Those inequalities yield an inequality which is opposite
    to the conjectured generalized Penrose inequality. They can thus
    be used to prove a uniqueness theorem for the \gK\ black holes if
    the generalized Penrose inequality can be established.}
\end{abstract}
\eject    
\tableofcontents

\section{Introduction}
Consider the families of metrics 
\begin{eqnarray}
  ds^2 & = & -(k - \frac {2m}r - \frac \Lambda 3 r^2) dt^2 + (k - \frac
  {2m}r - \frac \Lambda 3 r^2)^{-1} dr^2 + r^2 d\Omegak^2\ , \quad k =
  0, \pm 1\;, \nonumber\\  & & \label{Kot} \\
 ds^2 & = & -(\lambda - \Lambda  r^2) dt^2 + 
  (\lambda - \Lambda  r^2)^{-1} dr^2 + |\Lambda|^{-1} d\Omegak^2\ , \quad 
  k = \pm 1\;, \quad k\Lambda > 0\;, \lambda\in\R \nonumber\\ \label{Nar}
\end{eqnarray}
where $d\Omegak^2$ denotes a metric of constant Gauss curvature $k$ on
a two dimensional manifold ${}^2M$. (Throughout this work we assume
that ${}^2M$ is compact.) These are well known static solutions of the vacuum 
Einstein equation with a cosmological constant $\Lambda$; some subclasses 
of \eq{Kot} and \eq{Nar} have been
discovered by de Sitter \cite{deSitter1917b} (\eq{Kot} with $m = 0$
and $k=1$), by Kottler
\cite{Kottler} (Equation~\eq{Kot} with an arbitrary $m$ and $k=1$),
and by Nariai \cite{Nariai} (Equation~\eq{Nar} with $k=1$).  As
discussed in detail in Section \ref{Sghm}, the parameter $m\in \R$ is
related to the \GHn\ mass of the foliation $t=\mbox{const}$,
$r=\mbox{const}$. We will refer to those solutions as the \gK\ and the
generalized Nariai solutions.  The constant $\Lambda$ is an arbitrary
real number, but in this paper we will mostly be interested in
$\Lambda <0$, and this assumption will be made unless explicitly
stated otherwise.  There has been recently renewed interest in the
black hole aspects of the \gK\ 
solutions~\cite{GSWW,Vanzo,Manncqg,BLP}.  The object of this paper is
to initiate a systematic study of static solutions of the vacuum
Einstein equations with a negative cosmological constant.

   The first question that arises here is that of asymptotic conditions
   one wants to impose. In the present paper we consider metrics which 
   tend to the generalized Kottler  solutions, leaving the asymptotically 
   Nariai case to future work.   We present the following three
    approaches to asymptotic structure, and study their mutual
  relationships: three dimensional conformal compactifications, 
   four dimensional conformal completions, and a coordinate approach. 
  We show that under rather natural hypotheses the conformal boundary at 
  infinity is connected.
  
  The next question we address is that of the definition of mass for
  such solutions, \emph{without assuming staticity} of the metrics. We
  review again the possible approaches that occur here: a naive
  coordinate approach, a Hamiltonian approach, a ``Komar--type''
  approach, and the \GHn\ approach.  We
  show that the \GHn\ mass converges to a finite value for the metrics
  considered here, and we also give conditions on the conformal
  completions under which the ``coordinate mass'', or the Hamiltonian
  mass, are finite. Each of those masses come with different
  normalization factor, whenever all are defined, except for the Komar 
  and Hamiltonian masses which coincide. We suggest that the
  correct normalization is the Hamiltonian one.

Returning to the static case, we recall that appropriately behaved
vacuum black holes with $\Lambda = 0$ are completely described by the
parameter $m$ appearing above
\cite{Israel:vacuum,bunting:masood,Chstatic}, and it is natural to
enquire whether this remains true for other values of $\Lambda$. In
fact, for $\Lambda < 0$, Boucher, Gibbons, and Horowitz~\cite{BGH}
have given arguments suggesting uniqueness of the anti-de Sitter
solution within an appropriate class. As a step towards a proof of a 
uniqueness theorem in
the general case we derive, under appropriate hypotheses, 1) lower
bounds on (loosely speaking) the area of cross-sections of the
horizon, and 2) upper bounds on the mass of static vacuum black holes
with negative cosmological constant. When these inequalities are
combined the result goes precisely the opposite way as a (conjectured)
generalization of the Geroch--Huisken--Ilmanen--Penrose inequality
\cite{PenroseSCC,Geroch:extraction,HI1,HI2,Bray:thesis,Bray:preparation2}
appropriate to space-times with non-vanishing cosmological constant.
In fact, such a generalization was obtained by Gibbons~\cite{GibbonsGPI} along the lines of Geroch~\cite{Geroch:extraction}, 
and of Jang and Wald~\cite{JangWald}, {\em i.e.} 
under the very stringent assumption of the global existence
and smoothness of the inverse mean curvature flow, see
Section~\ref{GPI}. We note that it is far from clear that the
arguments of Huisken and Ilmanen~\cite{HI1,HI2}, or those of Bray
\cite{Bray:thesis,Bray:preparation2}, which establish the original
Penrose conjecture can be adapted to the situation at hand. If this
were the case, a combination of this with the results of the present
work would give a fairly general uniqueness result. In any case this
part of our work demonstrates the usefulness of a generalized Penrose
inequality, if it can be established at all.

To formulate our results more precisely, consider a static space-time
$(M,{}^4\!g)$ which might --- but does not have to --- contain a black
hole region.  In the asymptotically flat case there exists a well
established theory (see \cite{CarterHI}, or~\cite[Sections 2 and
6]{Chstatic} and references therein) which, under appropriate
hypotheses, allows one to reduce the study of such space-times to the
problem of finding all suitable triples $(\Sigma,g,V)$, where
$(\Sigma,g)$ is a three dimensional Riemannian manifold and $V$ is a
\emph{non-negative} function on $\Sigma$.  Further $V$ is required to
vanish precisely on the boundary of $\Sigma$, when non-empty:
\begin{equation}
  \label{f0}
  V\ge 0\;, \qquad V(p)=0 \Longleftrightarrow p\in \pSo \;.
\end{equation}
Finally $g$ and $V$ satisfy the following set of 
  equations 
on $\Sigma$:
\begin{eqnarray}
\label{f1}
\Delta V & = & - \Lambda V\ , \\
\label{f2}
R_{ij} & = & V^{-1} 
D_{i} D_{j} V + \Lambda  g_{ij}
\end {eqnarray}
($\Lambda=0$ in the asymptotically flat case). Here $R_{ij}$ is the
Ricci tensor of the (``three dimensional'') metric $g$.  We shall not
attempt to formulate the conditions on $(M,{}^4\!g)$ which will allow
one to perform such a reduction (some of the aspects of the
relationship between $(\Sigma,g,V)$ and the associated space-time are
discussed in Section~\ref{Sa4}, see in particular Equation
\eq{m1}), but we shall directly address the question of properties of solutions of
\eq{f1}--\eq{f2}. Our first main result concerns the topology of
$\pSo$ (\emph{cf.\/} Theorem~\ref{Tconn}, Section~\ref{Sconn}; compare
\cite{WittenYau,GSWW2}):

\begin{Theorem}
   \label{Ttop}
   Let $\Lambda<0$, consider a set $(\Sigma,g,V)$ which is $C^3$
   conformally compactifiable in the sense of Definition~\ref{Dc}
   below, suppose that \eq{f0}--\eq{f2} hold. Then the conformal
   boundary at infinity $\pSinfty$ of $\Sigma$ is connected.
\end{Theorem}

Our second main result concerns the \GHn\ mass of the level sets of
$V$, \emph{cf.\/} Theorem~\ref{Tmgh}, Section~\ref{Sghm}:

\begin{Theorem}
   \label{Tmass}
   Under the conditions of Theorem \ref{Ttop}, the \GHn\ mass $m$ of
   the level sets of $V$ is well defined and finite.
\end{Theorem}

It is natural to enquire whether there exist static vacuum
space-times with complete spacelike hypersurfaces and no black hole
regions; it is expected that no such solutions exist when $\Lambda <0$
and $\pSinfty\ne S^2$.  We hope that points\emph{~\ref{p2t1}}.~and
\emph{\ref{p3t1}}.~of the following theorem can be used as a tool to prove
their non-existence:

\begin{Theorem}
   \label{T1.1}
   Under the conditions of Theorem~\ref{Ttop}, suppose further that
   $$\pSo = \emptyset\;,$$ and that
   the scalar curvature $R'$ of the metric $g'=V^{-2}g$ is constant on
   $\pSinfty$. Then:
   \begin{enumerate} 
   \item \label{p1t1}If $\pSinfty$ is a sphere, then the \GHn\ mass
     $m$ of the level sets of $V$ is \emph{non-positive}, vanishing
     if and only if there exists a diffeomorphism
     $\psi:\Sigma\to\Sigma_0$ and a positive constant $\lambda$ such
     that $g=\psi^*g_0$ and $V=\lambda V_0\circ \psi$, with
     $(\Sigma_0,g_0,V_0)$ corresponding to the \mK\ space-time.
\item \label{p2t1}If $\pSinfty$ is a torus, then the \GHn\ mass $m$ is
  \emph{strictly negative}.
\item \label{p3t1}If the genus $g_\infty$ of $\pSinfty$ is higher than or equal to
  $2$ we have
   \begin{equation}
     \label{massineq}
     m< -\frac{1}{3\sqrt{-\Lambda}}\; ,
   \end{equation}
   with $m=m(V)$ normalized as in Equation \eq{masspar}.
 \end{enumerate}
\end{Theorem}

A mass inequality similar to that in point~\emph{\ref{p1t1}}.~above has been
established in \cite{BGH}, and in fact we follow their technique of
proof. However, our hypotheses are rather different. Further, the mass
here is \emph{a priori} different from the one considered in
\cite{BGH}; in particular it isn't clear at all whether the mass
defined as in \cite{BGH} is also defined for the metrics we consider,
\emph{cf.\/} Sections~\ref{Sa4c} and \ref{Scoord} below.

We note that metrics satisfying the hypotheses of point \ref{p2t1}
above, with arbitrarily large (strictly) negative mass, have been
constructed in \cite{HorowitzMyers}.

As a straightforward  corollary of Theorem \ref{T1.1} one has:

\begin{Corollary}
  \label{C1.1}
  Suppose that the generalized positive energy inequality $$m\ge
  m_{\mbox{\scriptsize crit}}(g_\infty)$$
  holds in the class of three
  dimensional manifolds $(\Sigma,g)$ which satisfy the requirements of
  point~\emph{1}.~of Definition~\ref{Dc} with a connnected conformal
  infinity $\pSinfty$ of genus $g_\infty$, and, moreover, the scalar
  curvature $R$ of which satisfies $R\ge2\Lambda$.  Then:
  \begin{enumerate}
  \item If $m_{\mbox{\scriptsize crit}}(g_\infty=0)=0$, then the only
    solution of Equations~\eq{f1}--\eq{f2} satisfying the hypotheses
    of point~\emph{\ref{p1t1}}.~of Theorem~\ref{T1.1} is the one
    obtained from anti-de Sitter space-time.
\item If $m_{\mbox{\scriptsize crit}}(g_\infty>1)=
  -1/(3\sqrt{-\Lambda})$, then there exist no solutions of
  Equations~\eq{f1}--\eq{f2} satisfying the hypotheses of
  point~\emph{\ref{p3t1}}.~of Theorem~\ref{T1.1}.
  \end{enumerate}
\end{Corollary}

When $\pSinfty=S^2$ one expects that the inequality $m\ge 0$,
\emph{with $m$ being the mass defined by spinorial identities} can be
established using Witten type techniques (\emph{cf.}~\cite{AndDahl,GHHP}),
regardless of whether or not $\pSo=\emptyset$. (On the other hand it
follows from \cite{Baum} that when $\pSinfty\ne S^2$ there exist no
asymptotically covariantly constant spinors which can be used in the
Witten argument.) This might require imposing some further restrictions
on \emph{e.g.} the asymptotic behavior of the metric. To be able to
conclude in this case that there are no static solutions without
horizons, or that the only solution with a connected non-degenerate
horizon is the \mK\ one, 
requires working out those restrictions,
and showing that the \GHn\ mass of the level sets of $V$ coincides
with the mass occuring in the positive energy theorem. 

When horizons occur, our comparison results for mass and area read as
follows:

\begin{Theorem}
   \label{T1.2}
   Under the conditions of Theorem~\ref{Ttop}, suppose further that
   the genus $g_\infty$ of $\pSinfty$ satisfies
$$g_\infty \ge 2\;, $$
and that the scalar curvature $R'$ of the metric $g'=V^{-2}g$ is
constant on $\pSinfty$. Let $\pSm$ be any connected component of $\pSo
$ for which the surface gravity $\kappa$ defined by
Equation~\eq{surfacegravity} is largest, and assume that
   \begin{equation}
     \label{kapineq}
   0 < \kappa \le \sqrt{-\frac {\Lambda}3}\ .
   \end{equation}
   Let $m_0$, respectively $A_0$, be the \GHn\ mass, respectively the
   area of $\pSo _0$, for that \gK\ solution $(\Sigma_0,g_0,V_0)$,
   with the same genus $g_\infty$, the surface gravity $\kappa_0$ of
   which equals $\kappa$. Then
 \begin{equation}
   \label{ourineq}
   m\le m_0\ , \qquad A_0(g_{\pSm} -1)\le A(g_{\infty}-1)\ ,
 \end{equation}
 where $A$ is the area of $\pSm $ and $m=m(V)$ is the \GHn\ mass of
 the level sets of $V$.  Further $m= m_0$ if and only if there exists 
a diffeomorphism  $\psi:\Sigma\to\Sigma_0$ and a positive constant 
$\lambda$ such that  $g=\psi^*g_0$ and $V=\lambda V_0\circ \psi$.
\end{Theorem}

 The asymptotic conditions assumed in Theorems~\ref{T1.1} and \ref{T1.2} are somewhat
related to those of~\cite{HT,HdS,AshtekarMagnonAdS,BGH}. The precise
relationships are discussed in Sections~\ref{Sa4} and~\ref{Sa4c}. Let
us simply mention here that the condition that $R'$ is constant on
$\pSinfty$ is 
the (local) higher genus analogue of the (global) condition in
\cite{HdS,AshtekarMagnonAdS} that the group of conformal isometries of
$\scri$ coincides with that of the standard conformal completion of
the \mK\ space-time; the reader is referred to
Proposition~\ref{Pconf} in Section~\ref{Sa4} for a precise statement.

We note that the hypothesis \eq{kapineq} is equivalent to the
assumption that the \gK\ solution with the same value of $\kappa$ has
non-positive mass; \emph{cf.\/} Section~\ref{SgK} for a discussion.
We emphasize, however, that we do not make any \emph{a priori}
assumptions concerning the sign of the mass of $(\Sigma,g,V)$. Our
methods do not lead to any conclusions for those values of $\kappa$
which correspond to \gK\ solutions with positive mass.

With $m=m(V)$ normalized as in Equation \eq{masspar}, the inequality $m \le
m_0$ takes the following explicit form
\begin{equation}
\label{Mle}
m \le 
\frac{(\Lambda + 2\kappa^2)\sqrt{\kappa^{2} - \Lambda}
  +2\kappa^3}{3\Lambda^{2}} \;,
\end{equation}
while $ A(g_\infty -1)\ge A_0(g_{\pSm} -1)$ 
can be explicitly written as 
\begin{equation}
\label{Rge}
A(g_\infty -1) \ge 4\pi(g_{\pSm} -1)\left[\frac{\kappa + \sqrt{\kappa^{2} -
      \Lambda}}{\Lambda}\right]^2 \;.
\end{equation}
(The right-hand sides of Equations~\eq{Mle} and \eq{Rge} are obtained
by straightforward algebraic manipulations from \eq{roots} and
\eq{surf}.)

It should be pointed out that in \cite{Woolgar:area} a lower bound for
the area has also been established.  However, while the bound there is sharp
only for the  generalized Kottler solutions with $m=0$, our bound is sharp for 
all Kottler solutions. On the other hand in
\cite{Woolgar:area} it is not assumed that the space-time is static.

If the generalized Penrose inequality  (which we discuss in some
detail in Section~\ref{GPI}) holds,
\begin{equation}\label{rdso1}
2  \mgh(u) \ge \sum_{i=1}^k
  \left((1-g_{\pSi}) \left(\frac{A_{\pSi}}{4\pi}\right)^{1/2}- \frac
  \Lambda 3  \left(\frac{A_{\pSi}}{4\pi}\right)^{3/2}\right) 
 \end{equation}
(with the $\pSi$'s, $i=1,\ldots,k$, being the connected
components of $\pSo$, the $A_{\pSi }$'s ---
their areas, and the $g_{\pSi }$'s --- the genera thereof) we obtain
uniqueness of solutions:

\begin{Corollary} 
  \label{C1.2}
  Suppose that the generalized Penrose inequality \eq{rdso1} holds in
  the class of three dimensional manifolds $(\Sigma,g)$  with scalar
  curvature $R$ satisfying $R\ge2\Lambda$,  which satisfy the
requirements of point~\emph{1}.~of 
  Definition~\ref{Dc} with  a connnected conformal infinity $\pSinfty$
of genus $g_\infty>1$, and which have a compact connected
  boundary. Then the only static solutions of Equations
  \eq{f1}--\eq{f2} satisfying the hypotheses of Theorem~\ref{T1.2} are
  the corresponding \gK\ solutions.
\end{Corollary}

This paper is organized as follows: in Section~\ref{SgK} we discuss
those aspects of the \gK\ solutions which are relevant to our work.
 The main object of Section~\ref{Sasymptotics} is to set forth
the boundary conditions which are appropriate for the problem at hand.
In Section~\ref{Sa3} this is analyzed from a three dimensional point
of view. We introduce the class of objects considered in Definition
\ref{Dc}, and analyze the consequences of this Definition in the
remainder of that section. In Section~\ref{Sa4} four-dimensional
conformal completions are considered; in particular we show how the
setup of Section~\ref{Sa3} relates to a four dimensional one,
\emph{cf.\/} Proposition~\ref{P34conf} and Theorem~\ref{T4conf}. We
also show there how the requirement of local conformal flatness of the
geometry of $\scri$ relates to the restrictions on the geometry of
$\pSinfty$ considered in Section~\ref{Sa3}. In Section~\ref{Sa4c} a
four dimensional coordinate approach is described; in particular, when
$(M,g)$ admits suitable conformal completions, we show there how to
construct useful coordinate systems in a neighborhood of $\scri$ ---
\emph{cf.\/} Proposition~\ref{Pcoord}. In Section~\ref{Sconn}
connectedness of the conformal boundary $\pSinfty$ is proved under
suitable conditions.  Section~\ref{Smass} is devoted to the question
how to define the total mass for the class of space-times at hand.
This is discussed from a coordinate point of view in Section~\ref{Scoord}, 
from a Hamiltonian point of view in  Section~\ref{Shamil}, and using the 
Hawking approach in Section~\ref{Sghm}; in
Section~\ref{SKomar} we present a generalization of the Komar integral 
appropriate to our setting.  The main results  of the analysis in 
Section~\ref{Smass} are the boundary conditions \eq{egk} together with 
Equation \eq{massequation}, which gives an ADM-type expression for 
the Hamiltonian mass for space-times with \gK\ asymptotics; we
emphasize that this formula holds without any hypotheses of staticity
or stationarity of the space-time metric.  Theorem~\ref{Tmass} is
proved in Section~\ref{Sghm}.  In Section~\ref{GPI} we recall an
argument due to Gibbons \cite{GibbonsGPI} for the validity of the
generalized Penrose inequality. (However, our conclusions are
different from those of \cite{GibbonsGPI}.) In Section~\ref{spT1} we
prove Theorems~\ref{T1.1} and \ref{T1.2}, as well as Corollary
\ref{C1.2}.

\textbf{Acknowledgements} W.S. is grateful to Tom Ilmanen for helpful
discussions on the Penrose inequality. We  thank Gary Horowitz for
pointing out Reference~\cite{HorowitzMyers}.
 
\section{The \gK\ solutions}
\label{SgK}

We recall some properties of the solutions \eq{Kot}.  Those solutions
will be used as reference solutions in our arguments, so it is
convenient to use a subscript $0$ when referring to them. As already
mentioned, we assume $\Lambda<0$ unless indicated otherwise.
For $m_0\in \R$, let $r_0$ be the largest positive root of the
equation
\footnote{See~\cite{Vanzo} for an exhaustive analysis, and
  explicit formulae for the roots of Equation~\eq{roots}.}
\begin{equation}
  \label{roots}
  V_0^2\equiv k - \frac {2m_0}r - \frac \Lambda 3 r^2 = 0\ .
\end{equation}
We set 
\begin{eqnarray}
  \label{refm}
 & \Sigma_0=\{(r,v)|r> r_0, v\in {}^2M\}\;, &
\\  &\label{refg}\displaystyle g_0= (k - \frac
  {2m_0}r - \frac \Lambda 3 r^2)^{-1} dr^2 + r^2 d\Omegak^2\ , &
\end{eqnarray}
where, as before, $d\Omegak^2$ denotes a metric of constant Gauss
curvature $k$ on a smooth two dimensional compact manifold ${}^2M$. We
denote the corresponding surface gravity by $\kappa_0$. (Recall that
the surface gravity of a connected component of a horizon $\cNX$ is
usually defined by the equation
\begin{equation}
  \label{kdef}
  (X^\alpha X_\alpha)_{,\mu}\Big|_{\cNX} = -2\kappa X_\mu \ ,
\end{equation}
where $X$ is the Killing vector field which is tangent to the
generators of $\cNX$. This requires normalizing $X$; here we impose
the normalization\footnote{When $\Mtwo=T^2$ a unique normalization of
  $X$ needs a further normalization of $d\Omegak^2$, \emph{cf.\/}
  Sections~\ref{Scoord} and \ref{Shamil} for a detailed discussion of
  this point.} that $X=\partial/\partial t$ in the coordinate system
of \eq{Kot}.)  We set
\begin{equation}
\label{W0}
W_{0}(r)  \equiv  g_{0}^{ij} D_{i}V_{0} D_{j}V_{0} 
= (\frac{m_0}{r^{2}} - \frac{\Lambda r}{3})^{2}\ .
\end{equation}
When $m_0=0$ we note the relationship
\begin{equation}
  \label{krel}
  W_0=  - \frac{\Lambda }{3}(V^2_{0}-k)\;,
\end{equation}
which will be useful later on, and which holds regardless of the
topology of  ${}^2M$.
\subsection{ $k=-1$}
Suppose, now, that $k=-1$, and that $m_0$ is in the
range
\begin{equation}
\label{rngm}
m_0 \in [m_{\mbox{\scriptsize crit}},0] \ ,
\end{equation} where
\begin{equation}
  \label{mcrit}
m_{\mbox{\scriptsize crit}}\equiv -\frac{1}{3\sqrt{-\Lambda}}\ .  
\end{equation}
Here $m_{\mbox{\scriptsize crit}}$ is defined as the smallest value of
$m_0$ for which the metrics \eq{Kot} can be extended across a Killing
horizon~\cite{Vanzo,BLP}.  
Let us show that Equation~\eq{rngm} is equivalent to 
\begin{equation}\label{rngr} r_0
  \in [\frac{1}{\sqrt{-\Lambda}}, \sqrt{-\frac{3}{\Lambda}}]\;.
\end{equation}
In order to simplify notation it is useful to introduce
\begin{equation}\label{Lshort}
  \frac 1{\ell^2}\equiv -\frac \Lambda 3\ .
\end{equation}
Now, the equation $V_0(\ell/\sqrt{3})=0$ implies
$m=m_{\mbox{\scriptsize crit}}$. Next, an elementary analysis of the
function $r^3/\ell^2-r-2m_0$ (recall that $k=-1$ in this section)
shows that 1) $V$ has no positive roots for $m<m_{\mbox{\scriptsize
    crit}}$; 2) for $m=m_{\mbox{\scriptsize crit}}$ the only positive
root is $\ell/\sqrt 3$; 3) if $r_0$ is the largest positive root of
the equation $V_0(r_0)=0$, then for each $m_0>m_{\mbox{\scriptsize
    crit}}$ the radius $r_0(m_0)$ exists and is a differentiable
function of $m_0$. Differentiating the equation $r_0V_0(r_0)=0$ with
respect to $m_0$ gives $(\frac{3r_0^2}{\ell^2}+k)\frac{\partial
  r_0}{\partial m_0}=(\frac{3r_0^2}{\ell^2}-1)\frac{\partial
  r_0}{\partial m_0}=2$. It follows that for $r\ge\ell/\sqrt{3}$ the
function $r_0(m_0)$ is a monotonically increasing function on its
domain of definition $[m_{\mbox{\scriptsize crit}},\infty)$, which
establishes our claim. 

We note that the surface gravity $\kappa_0$ is given by the formula
\begin{equation}
  \label{surf}
  \kappa_0 = \sqrt{W_0(r_0)}= \frac {m_0} {r_0^2} + \frac
  {r_0}{\ell^2}\ ,
\end{equation}
which gives 
$$\frac {\partial \kappa_0}{\partial m_0}= \frac 1 {r_0^2} +
\left(\frac {1}{\ell^2} - \frac {2m_0} {r_0^3}\right) \frac{\partial
  r_0}{\partial m_0}\ .  $$ Equation~\eq{surf} shows that $\kappa_0$
vanishes when $m_0=m_{\mbox{\scriptsize crit}}$.
\footnote{The methods of~\cite{Walker} show that in this case
  the space-times with metrics \eq{Kot} can be extended to black hole
  space-times with a degenerate event horizon, thus a claim to the
  contrary in~\cite{Vanzo} is wrong. It has been claimed without proof 
  in~\cite{BLP} 
  that   $\scrp$, as constructed by the methods of~\cite{Walker}, can be
  extended to a larger one, say $\widehat{\scrp}$, which is
  connected. 
  Recall that that claim would imply that $\partial
  I^{-}(\widehat{\scrp})=\emptyset$ (see Figure~2 in~\cite{BLP}), thus
  the space-time would not contain an event horizon with respect to
  $\widehat{\scrp}$. Regardless of whether such an extended $\widehat{\scrp}$ 
  exists or not, we wish to point out the following:
  a) there will still be degenerate \emph{event} horizons as defined
  with respect to any connected component of $\scrp$; b) regardless of 
  how null infinity is added 
  there will exist \emph{degenerate Killing horizons} in those
  space-times; c) there will exist an \emph{observer} horizon
  associated to the world-line of any observer which moves along the
  orbits of the Killing vector field in the asymptotic region. It thus
  appears reasonable to give those space-times a black hole
  interpretation in any case.} {\em \/Under the hypothesis that}
$m_0\le0$, it follows from 
what has been said above a) that $\frac {\partial
  \kappa_0}{\partial m_0}$ is positive; b) that we have
\begin{equation}
\label{rngk}
\kappa_0 \in [0, \sqrt{-\frac{\Lambda}{3}}] \; ,
\end{equation}
when \eq{rngm} holds, and c) that, under the current hypotheses on $k$
and $\Lambda$, \eq{rngm} is equivalent to \eq{rngk} for the metrics
\eq{Kot}. While this can probably be established directly, we note
that it follows from Theorem~\ref{T1.2} that \eq{rngk} is equivalent to
\eq{rngm} without having to assume that $m_0\le 0$.

In what follows we shall need the fact that in the above ranges of
parameters the relationship $V_0(r)$ can be inverted to define a
smooth function $r(V_0):[0,\infty)\to\R$. Indeed, the equation
$\frac{dV_0}{dr}(r_{\mbox{\scriptsize{crit}}})=0$ yields
$r^3_{\mbox{\scriptsize{crit}}}=3m_0/\Lambda$; when $k=-1$, $\Lambda <
0$, and when \eq{rngm} holds one finds
$V_0(r_{\mbox{\scriptsize{crit}}})\le0$, with the inequality being
strict unless $m=m_{\mbox{\scriptsize{crit}}}$.  Therefore, $V_0(r)$
is a smooth strictly monotonic function in $[r_0, \infty)$, which
implies in turn that $r(V_0)$ is a smooth strictly monotonic function
on $(0,\infty)$; further $r(V_0)$ is smooth up to $0$ except when
$m=m_{\mbox{\scriptsize{crit}}}$.

\section{Asymptotics}\label{Sasymptotics}

\subsection{Three dimensional formalism}\label{Sa3}

As a motivation for the definition below, consider one of the metrics
\eq{Kot} and introduce a new coordinate $x \in (0, x_0]$ by
\begin{equation}
  \label{defr}
  \frac{r^2}{\ell^2} = \frac{1-kx^2}{x^2}\;.
\end{equation}
with  $x_0$ defined by substituting $r_0$ in the left-hand-side of \eq{defr}.
It then follows that
$$g= \ell^2 x^{-2} \left[\frac{dx^2}{(1 - kx^2)
(1 - \frac{2mx^3}{\ell\sqrt{1 - kx^2}})} +
(1 - kx^2)d\Omega_{k}^{2}\right]\;.$$
Thus the metric
$$g'\equiv (\ell^{-2}x^2)g$$ is a smooth up to boundary metric on the compact 
manifold with boundary $\bSo\equiv [0, x_0] \times {}^2M$. Furthermore, 
$x V_0$ can be extended by continuity to a smooth up to boundary function on 
$\bSo$, with $x V_0 = 1$. This justifies the following definition:

\begin{Definition}
  \label{Dc} Let $\Sigma$ be a smooth manifold\footnote{All manifolds
    are assumed to be Hausdorff, 
    paracompact, and orientable throughout.}, with perhaps a compact
  boundary which we denote by $\pSo $ when non empty.\footnote{We use
    the convention that a manifold with boundary $\Sigma$ contains its
    boundary as a point set.} Suppose that $g$ is a smooth metric on
  $\Sigma$, and that $V$ is a smooth nonnegative function on $\Sigma$,
  with $V(p)=0$ if and only if $p\in\pSo $.
\begin{enumerate}
\item $(\Sigma,g)$ will be said to be $C^\kind$, $\kind\in
  \N\cup\{\infty\}$, conformally compactifiable or, shortly, compactifiable, if
 \label{first} there exists a $C^{\kind+1} $ diffeomorphism $\chi$
 from $\Sigma\setminus\pSo $ to the interior of a compact Riemannian
 manifold with boundary $(\bS\approx \Sigma\cup\pSinfty,\bg)$, with
 $\pSinfty\cap\Sigma =\emptyset$, and a $C^\kind$ function
 $\varomega:\bS\to \R^+$ such that
    \begin{equation}
      \label{chidiff}
      g=\chi^*(\varomega^{-2}\bg)\;.
    \end{equation}
 We further assume that
  $\{\varomega=0\}=\pSinfty$, with $d\varomega$ nowhere vanishing on
  $\pSinfty$,  and that $\bg$ is of $C^{i}$ differentiability class on $\bS$.
\item \label{second} A triple $(\Sigma,g,V)$ will be said to be
  $C^\kind$, $\kind\in \N\cup\{\infty\}$, compactifiable if
  $(\Sigma,g)$ is $C^\kind$ compactifiable, and if $V\varomega$
  extends by continuity to a $C^\kind$ function on $\bS$,
\item \label{third}with
  \begin{equation}
    \label{pos}
\lim_{\varomega\to  0}V \varomega> 0\;.
  \end{equation}
  \end{enumerate}
\end{Definition}

We emphasize that $\Sigma$ itself is allowed to have a boundary on
which 
$V$  vanishes, 
$$\pSo =\{p\in\Sigma|
V(p)=0\}\;.$$ If that is the case we will
have $$\partial\bS= \pSo \cup\pSinfty\;.$$ 

To avoid ambiguities, we stress that one point compactifications of the kind encountered in the
asymptotically flat case (\emph{cf., e.g.,}~\cite{BeigSimon}) are not
allowed in our context. 

The conditions above are not
independent when the ``static field equations''
(Equations~\eq{f1}--\eq{f2}) hold:
\begin{Proposition}
  \label{Pnotind}
  Consider a triple $(\Sigma,g,V)$ satisfying Equations
  \eq{f0}--\eq{f2}.
\begin{enumerate}
\item The condition that $|d\varomega|_{\bg}$ has no zeros on $\pSinfty$
follows from the remaining hypotheses of point\emph{~\ref{first}}.~of
Definition~\ref{Dc}, when those hold with $\kind \ge 2$.
\item Suppose that $(\Sigma,g)$ is $C^\kind$ compactifiable with
  $\kind \ge 2$. Then $\lim_{\varomega\to 0}V \varomega$ exists.
  Further, one can choose a (uniquely defined) conformal factor so
  that $\varomega$ is the $\bg$-distance from $\pSinfty$. With this
  choice of conformal factor, when \eq{pos} holds a necessary
  condition that $(\Sigma,g,V)$ is $C^\kind$ compactifiable is that
  \be\label{restr}
(4{ \bR}_{ij} - { \bR}\bg_{ij}) \bn^{i}\bn^{j} 
\Big|_{\pSinfty}=0\;, \ee where $\bn$ is the 
field of unit normals to $\pSinfty$.
\item 
  $(\Sigma,g,V)$ is $C^\infty$ compactifiable if and only if
  $(\Sigma,g)$ is $C^\infty$ compactifiable and Equations \eq{pos} and
  \eq{restr} hold.
\end{enumerate}
\end{Proposition}

\noindent {\sc Remarks:} 1.  When $(\Sigma,g)$ is $C^\infty$
compactifiable but Equation~\eq{restr} does not hold, the proof 
below shows that
$V\varomega$ is of the form $\alpha_0 + \alpha_1
\varomega^2\log\varomega$, for some smooth up-to-boundary functions
$\alpha_0$ and $\alpha_1$.
This isn't perhaps so surprising because the nature of the equations
satisfied by $g$ and $V$ suggests that both $\bg$ and $V\varomega$
should be polyhomogeneous, rather than smooth.  (``Polyhomogeneous''
means that $\bg$ and $V\varomega$ are expected to admit asymptotic
expansions in terms of powers of $\varomega$ and $\log \varomega$ near
$\pSinfty$ under some fairly weak conditions on their behavior at
$\pSinfty$; \emph{ cf., e.g.\/} \cite{AndChDiss} for precise
definitions and related results.) {}From this point of view the
hypothesis that $(\Sigma,g)$ is $C^\infty$ compactifiable is somewhat
unnatural and should be replaced by that of polyhomogeneity of $\bg$
at $\pSinfty$.

2. One can prove appropriate versions of point~\emph{3}.~above for
$(\Sigma,g)$'s which are $C^\kind$ compactifiable for finite $\kind$.
This seems to lead to lower differentiability of $1/V$ near $\pSinfty$
as compared to $\bg$, and for this reason we shall not discuss it
here.

3. We leave it as an open problem whether or not there exist solutions
of \eq{f0}--\eq{f2} such that $(\Sigma,g)$ is smoothly compactifiable,
such that $V$ can be extended by continuity to a smooth function on
$\bS$, while \eq{pos} does not hold.

4. We note that \eq{restr} is a conformally
invariant condition because $\omega$ and $\bg$ are uniquely determined
by $g$. However, it is not conformally covariant, in
the sense that if $\bg$ is conformally rescaled, then \eq{restr} will
not be of the same form in the new rescaled metric. It would be of
interest to find a form of \eq{restr} which does not have this
drawback.

5. The result above has counterparts for one-point compactifications
in the asymptotically flat case, ({\it cf., e.g.}, the theorem in the
Appendix of \cite{BeigSimon}).

\medskip

\proof Let
$$\alpha\equiv V\varomega\;.$$ After suitable identifications we can
without loss of generality assume that the map $\chi$ in \eq{chidiff}
is the identity. Equations
\eq{f1}--\eq{f2} together with the definition of 
$\bg=\varomega^{2}g$ lead to the following
\begin{eqnarray}
\label{a1.1}
& \bDelta \alpha - 3 
\frac{\bD^i\varomega \bD_i \alpha}\varomega +
\left(\frac{ \bDel \varomega}{\varomega} + \frac {\bR} 2 
 \right)\alpha = 0\;,
& \\ & \label{a2.1}
\bD_i \bD_j \alpha -
\frac{\bD^k\varomega \bD_k \alpha}\varomega \bg_{ij} = \left(\bR_{ij}
  +2\frac{\bD_i \bD_j\varomega}\varomega -\left(\frac{ \bDel
      \varomega}{\varomega} + \frac {\bR} 2 \right)
  \bg_{ij}\right)\alpha\;.
\end{eqnarray}
We have also used $R=2\Lambda$ which, together with the transformation
law of the curvature scalar under conformal transformations, implies
\begin{equation}
 \label{a0}
\varomega^{2}\bR =  6 |d\varomega|^2_{\bg} +2\Lambda - 4\varomega \bDel
\varomega\;. 
\end{equation}
In all the equations here barred quantities refer to the metric $\bg$.
Point \emph{1} of the proposition follows immediately from Equation~\eq{a0}.

To  avoid factors of $-\Lambda/3$ in the remainder of the
proof we rescale the metric $g$ so that $\Lambda = -3$. Next, to avoid
annoying technicalities we shall present the proof only for smoothly
compactifiable $(\Sigma,g)$ --- $\kind=\infty$; the finite $\kind$
cases can be handled using the results in \cite[Appendix A]{AndChDiss}
and~\cite[Appendix A]{ChDGH}.  Suppose, thus, that $\kind=\infty$. As
shown in~\cite[Lemma 2.1]{AChF} we can choose $\varomega$ and $\bg$ so
that $\varomega$ coincides with the $\bg$-distance from $\pSinfty$ in
a neighborhood of $\pSinfty$; we shall use the symbol $x$ to denote
this function. In this case we have
\begin{equation}
  \label{e2}
  \bDel \varomega = \bp\;,
\end{equation}
where $\bp$ is the mean curvature of the level sets of $\varomega=x$.
Further $|d\varomega|_{\bg}=1$ so that \eq{e2} together with \eq{a0}
give
\begin{equation}
  \label{e3}
  \bR = -4 \frac \bp x \;,
\end{equation}
in particular 
\begin{equation}
  \label{e4.0}
  \bp\Big|_{x=0} = 0\;.
\end{equation}
We can introduce Gauss coordinates $(x^1,x^A)$ near $\pSinfty$ in which
$x^1=x\in [0,x_0)$, while the $(x^A)=v$'s form local coordinates on $\pSinfty$,
with the metric taking the form
\begin{equation}
  \label{Gausscoord}
\bg= dx^2+\bh\ , \qquad \bh(\partial_x,\cdot)=0\;.
\end{equation}
To prove point~\emph{2}, from Equation~\eq{a2.1} we obtain
 \begin{eqnarray}
   \lefteqn{\varomega\bD^i\varomega \bD^j\varomega\bD_i(\varomega^{-1} \bD_j
   \alpha ) \; = }     \nonumber
\\ & = & \bD^i\varomega \bD^j\varomega \left(\bR_{ij}
  +2\frac{\bD_i \bD_j\varomega}\varomega -\left(\frac{ \bDel
      \varomega}{\varomega} + \frac {\bR} 2 \right)
  \bg_{ij}\right)\alpha\;.    \label{e0m}
 \end{eqnarray}
 Equations \eq{e2}--\eq{e0m} lead to
\begin{equation}
  \label{e4}
  x\partial_x(x^{-1}\partial_x \alpha) = (\bR_{xx}-\frac \bR 4 ) \alpha\;.
\end{equation}
At each $v\in \pSinfty$ this is an ODE of Fuchsian type for
$\alpha(x,v)$. Standard results about such equations show that for
each $v$ the functions $x\to\alpha(x,v)$ and
$x\to\partial_x\alpha(x,v)$ are bounded and continuous on $[0,x_0)$.
Integrating \eq{e4} one finds
\begin{equation}
  \label{e5}
  \partial_x \alpha = x\beta(v) + (\bR_{xx}-\frac \bR 4 )\alpha(0,v) x
  \ln x + O(x^2\ln x)\;,
\end{equation}
where $\beta(v)$ is a ($v$-dependent) integration
constant.
By hypothesis there exist no points at $\pSinfty$ such that $\alpha(0,v)=0$,
Equations \eq{e4} and \eq{e5} show that $\partial^2_x \alpha$ blows up
at $x=0$ unless \eq{restr} holds, and point~\emph{2}.~follows.

We shall only sketch the proof of point~\emph{3}.: Standard results about
Fuchsian equations show that solutions of Equation~\eq{e4} will be
smooth in $x$ whenever $(\bR_{xx}-\frac \bR 4 )(x=0,v)$ vanishes
throughout $\pSinfty$. A simple bootstrap argument applied to Equation
\eq{a2.1} with $(ij)=(1A)$ shows that $\alpha$ is also smooth in $v$.
Commuting Equation~\eq{a2.1} with $(x\partial_x)^i\partial_v^\beta$,
where $\beta$ is an arbitrary multi-index, and iteratively repeating
the reasoning outlined above establishes smoothness of $\alpha$
jointly in $v$ and $x$.  \qed

A consequence of condition~\ref{third} of Definition~\ref{Dc} is that
the function 
$$V'\equiv1/V\;,$$ when extended to $\bS$ by setting $V'=0$ on $\pSinfty$,
can be used as a compactifying conformal factor, at least away from
$\pSo $: If we set
$$g'= V^{-2} g\;,$$ then $g'$ is a Riemannian metric smooth up to
boundary on $\bS\setminus \pSo $. In terms of this metric Equations \eq{f1}--\eq{f2} can be rewritten as
\begin{eqnarray}
\label{f3}
\Delta' V' & = & 3 V'W+ \Lambda V\ , \\
\label{f4}
R'_{ij} & = & -2 V D'_{i} D'_{j} V' \;.
\end {eqnarray}
Here $R'_{ij}$ is the Ricci tensor of the metric $g'$, $D'$ is the
Levi--Civita covariant derivative associated with $g'$, while
$\Delta'$ is the Laplace operator associated with $g'$.  Taking the
trace of \eq{f4} and using \eq{f3} we obtain
\begin{equation}
  \label{Rprime}
  R'= -6 W - 2\Lambda V^2\;,
\end{equation}
where
\begin{equation}
  \label{Wdef}
W \equiv D_i V D^i V\;.
\end{equation}
Defining 
\begin{equation}
  \label{Wprime}
  W'\equiv g'^{ij}D'_iV'D'_jV'= (V')^2W\;, 
\end{equation}
Equation~\eq{Rprime} can be rewritten as
\begin{equation}
  \label{Rprime5}
    6W'= -2\Lambda -R'(V')^2\;.
\end{equation}
If $(\Sigma,g,V)$ is $C^2$ compactifiable then $R'$ is bounded in a
neighborhood of $\pSinfty$, and since $V$ blows up at $\pSinfty$ it follows from
Equation~\eq{Rprime} that so does $W$, in particular $W$ is strictly
positive in a neighborhood of $\pSinfty$. Further Equation~\eq{Rprime5}
implies that the level sets of $V$ are smooth manifolds in a
neighborhood of $\pSinfty$, diffeomorphic to $\pSinfty$ there.

Equations \eq{f1}--\eq{f2} are invariant under a rescaling
$V\to\lambda V$, $\lambda \in \R^*$. This is related to the
possibility of choosing freely the normalization of the Killing vector
field in the associated space-time. Similarly the conditions of
Definition~\ref{Dc} are invariant under such rescalings with $\lambda
>0$. For various purposes --- \emph{e.g.,}, for the definition
\eq{surfacegravity} of surface gravity --- it is convenient to have a
unique normalization of $V$. We note that if $(\Sigma,g,V)$
corresponds to a \gK\ solution $(\Sigma_0,g_0,V_0)$, then \eq{Kot} and
\eq{W0} together with \eq{Wdef} give $ 6W_0'= -2\Lambda(1 -k (V_0')^2)
+ O((V_0')^3)$ so that from  \eq{Rprime} one obtains
\begin{equation}
  \label{Rprime0.0}
  R_0'|_{\pSinfty}= -2\Lambda  k\;.
\end{equation}
We have the following:
\begin{Proposition}
  \label{P2}  Consider a $C^\kind$-compactifiable  triple
  $(\Sigma,g,V)$, $\kind\ge 3$, satisfying equations \eq{f1}--\eq{f2}.
  \begin{enumerate}
  \item We have 
\begin{equation}
  \label{RR}
  {}^{2}{\cal R'}\Big|_{x=0}=\frac 13 R'\Big|_{x=0}\;,
\end{equation}
where ${}^{2}{\cal R'}$ is the scalar curvature of the metric induced
by $g'\equiv V^{-2}g$ on the level sets of $V$, and $R'$ is the Ricci
scalar of $g'$.
\item If $R'$ is constant on ${\pSinfty}$, replacing  $V$ by a
  positive multiple thereof if necessary we can achieve
  \begin{equation}
  \label{Rprime0}
  R'|_{\pSinfty}= -2\Lambda  k\;,
\end{equation}
where $k=0$, $1$ or $-1$ according to the sign of the Gauss curvature
of the metric induced by $g'$ on $\pSinfty$.
  \end{enumerate}
\end{Proposition}

\noindent {\sc Remark:} When $k=0$ Equation~\eq{Rprime0} 
holds with an arbitrary normalization of $V$.
\medskip

\proof 
Consider a level set
$\{V= \mbox{const}\}$ of $V$ which is a smooth hypersurface in
$\overline{\Sigma}$, with unit normal $n_{i}$,  induced metric $h_{ij}$, scalar curvature 
${}^{2}{\cal R}$, second fundamental form $p_{ij}$ defined with
respect to an inner pointing normal, mean curvature $p =
h^{ij}p_{ij}={h_i}^k{h_j}^m D_{(k}n_{m)}$; we denote by $q_{ij}$ the
trace-free part of $p_{ij}$: $q_{ij} = p_{ij} - 1/2h_{ij}p$.  Let
$R_{ijk}$, respectively $R'_{ijk}$, be the Cotton tensor of the metric
$g_{ij}$, respectively $g'_{ij}$; by definition,
\begin{equation}
  \label{Cotton}
  R_{ijk}= 2 \left(R_{i[j}-\frac 14Rg_{i[j}\right){}_ {;k]}\;,
\end{equation}
where square brackets denote antisymmetrization with an appropriate
combinatorial factor ($1/2$ in the equation above), and a semi-column
denotes covariant differentiation. We note the useful
identity due to Lindblom \cite{LL} 
\begin{eqnarray}
  \label{LL}
  R'_{ijk} R'^{ijk} & = & V^6 R_{ijk}R^{ijk}\nonumber
\\ & = & 8 (VW)^2 q_{ij} q^{ij} +  V^2 h^{ij}D_i W D_j W\ .
\end{eqnarray}
When $(\Sigma,g,V)$ is $C^3$ compactifiable the function $ R'_{ijk}
R'^{ijk}$ is uniformly bounded on a neighborhood of $\bS$, which
gives
\begin{equation}
  \label{qfall}
   (VW)^2 q_{ij} q^{ij} \le C
\end{equation}
in that same neighborhood, for some constant $C$. Equations \eq{qfall}
 and \eq{Wprime} give
 \begin{equation}
   \label{qfall2}
   |q|_{g} = O((V')^3)\;,
 \end{equation}
 Let $q'_{ij}$ be the trace-free part of the second fundamental form
 $p'_{ij}$ of the level sets of $V'$ with respect to the metric
 $g'_{ij}$,  defined with respect to an inner pointing normal; we
 have $q'_{ij}= q_{ij}/V$, so that
 \begin{equation}
   \label{qfall3}
   |q'|_{g'} = O((V')^2)\;.
 \end{equation}
 Throughout we use $|\cdot|_{k}$ to denote the norm of a
 tensor field with respect to a metric $k$.

 Let us work out some implications of \eq{qfall3}; Equations
 \eq{f3}--\eq{Rprime} lead to
\begin{equation}
  \label{f5}
  (\Delta'+\frac {R'}2 ) V' = 0 \;.
\end{equation}
Equations \eq{Wprime} and \eq{Rprime5} show that $dV'$ is nowhere
vanishing on a suitable neighborhood of $\pSinfty$.  We can thus
introduce coordinates there so that
$$V'=x\;.$$
If the remaining coordinates are Lie dragged along the integral curves
of $D'x$ the metric takes the form
\begin{equation}
  \label{gprime}
 g'=(W')^{-1} \, dx^2+h'\;, \qquad h'(\partial_x,\cdot)=0\;.
\end{equation}
Equations \eq{f5}--\eq{gprime} give then
\begin{eqnarray}
  \label{pprime}
  \nonumber p' & = & -\frac 1 {2\sqrt{W'}}\left( \frac{\partial W'}{\partial
  x} + 
R'
x\right)
\\ & = & - \frac{x}{12\sqrt{W'}}\left( 4R'-x\frac{\partial R'}{\partial
  x}\right)\;,
\end{eqnarray}
and in the second step we have used \eq{Rprime5}.  Here $p'= \sqrt{W'}
\partial_x(\sqrt{\det h'})/\sqrt{\det h'}$ is the mean curvature of
the level sets of $x$ measured with respect to the inner pointing
normal $n'=\sqrt{W'}\partial_x$.
Equation~\eq{f4} implies
\begin{eqnarray*}
  R'_{ij}n'^i n'^j & = & -2V n'^i n'^j D'_i D'_jV'
\\ & = & - 2 \frac{D'^i V'D'^jV'}{V'W'}D'_i D'_jV'
\\ & = & -  \frac{D'^i V'D'_iW'}{V'W'} \;=\; \frac{-\partial_x W'}{x}
\end{eqnarray*}
in the coordinate system of Equation~\eq{gprime}. {}From \eq{Rprime5} we
get
\begin{equation}
  \label{Rnn}
  R'_{ij}n'^i n'^j  = \frac{R'}3 + O(x)\;.
\end{equation}
{}From the Codazzi--Mainardi equation,
\begin{eqnarray}
 (-2 { R'}_{ij} + { R'}g'_{ij}) n'^{i}n'^{j} &=& {}^{2}{\cal R'} +
q'_{ij}q'^{ij} - \frac{1}{2}p'^{2} 
\;,
  \label{CMnn}
\end{eqnarray}
where ${}^{2}{\cal R'}$ is the scalar curvature of the metric induced
by $g'$ on $\pSinfty$, one obtains
\begin{eqnarray}
 (-2 { R'}_{ij} + { R'}g'_{ij}) n'^{i}n'^{j} &= &  {}^{2}{\cal R'} +O(x) \;,
  \label{CMnn2}
\end{eqnarray}
where we have used \eq{qfall3} and \eq{pprime}. This, together with
Equation~\eq{Rnn}, establishes Equation~\eq{RR}. In particular
$R'|_{\pSinfty}$ is constant if and only if ${}^{2}{\cal R'}$ is, and
$R'$ at $x=0$ has the same sign as the Gauss curvature of the relevant
connected component of $\pSinfty$. Under a rescaling $V\to\lambda V$,
$\lambda >0$, we have  
$W\to\lambda^2 V$; Equation~\eq{Rprime} shows that $R'\to\lambda^2R'$,
and choosing $\lambda$ appropriately establishes the
result. \qed

We do not know whether or not there exist smoothly compactifiable
solutions of Equations \eq{f1}--\eq{f2} for which $R'$ is not locally
constant at $\pSinfty$, it would be of interest to settle this
question. Let us point out that the remaining Codazzi--Mainardi
equations do not lead to such a restriction. For example, consider the
following equation:
\begin{eqnarray}
  R'_{1a} & = & -\cD'_ap'+ \cD'_b{p'_a}^b \nonumber
\\ & = & -\frac 12 \cD'_ap'+ \cD'_b{q'_a}^b\;. \label{CM1a}
\end{eqnarray}
Here we are using the adapted coordinate system of Equation
\eq{gprime} with $x^1=x$ and with the indices $a,b=2,3$ corresponding
to the remaining coordinates; further $\cD'$ denotes the Levi--Civita
derivative associated with the metric $h'$. Since $D'_a D'_x x= \cD'_a
\sqrt{W'} $, Equation~\eq{f4} yields
\begin{equation}
  \label{f7}
  \cD'_a( \sqrt{W'}- \frac 14 xp') = -\frac 12 x \cD'_b{q'_a}^b= O(x^3)\;;
\end{equation}
in the last equality Equation~\eq{qfall3} has been used.
Unfortunately the terms containing $R'$ exactly cancel out in
Equations \eq{pprime} and \eq{Rprime5} leading to
$$\sqrt{W'}+ \frac 14 xp'= \sqrt{-\frac {\Lambda}{3}}
+O(x^3)\;,$$
which does not provide any new information.
\subsection{Four dimensional conformal approach}\label{Sa4}

Consider a space-time $(M,\fg)$  of the form $M=\R\times\Sigma$ with
the metric $\fg$ 
\begin{equation}
  \label{m1}
  \fg = -V^2dt^2 + g\;,\quad g(\partial_t,\cdot)=0\;, \quad \partial_tV
  = \partial_tg=0\;.
\end{equation}
By definition of a space-time $\fg$ has Lorentzian signature, which
implies that $g$ has signature $+3$; it then naturally defines a
Riemannian metric on $\Sigma$ which will still be denoted by $g$.
Equations \eq{f1}--\eq{f2} are precisely the vacuum Einstein equations
with cosmological constant $\Lambda$ for the metric $\fg$. It has been
suggested that an appropriate~\cite{HdS,AshtekarMagnonAdS} framework
for asymptotically anti-de Sitter space-times is that of conformal
completions introduced by Penrose \cite{penrose:scri}. The work of
Friedrich~\cite{Friedrich:adS} has confirmed that it is quite
reasonable to do that, by showing that a large class of space-times
(not necessarily stationary) with the required properties exist; some
further related results can be found in~\cite{Kannar:adS,Magnon}.  In
this approach one requires that there exists a space-time with
boundary $(\bM,\fbg)$ and a positive function $\varOmega:\bM\to\R^+$,
with $\varOmega$ vanishing precisely at $\pM\subset\partial \bM$, and
with $d\Omega$ without zeros on $\pM$, together with a diffeomorphism
$\Chi:M\to\bM\setminus\pM$ such that
\begin{equation}
  \label{m2}
  \fg = \Chi^{*}(\varOmega^{-2}\;\fbg)\;.
\end{equation}
The vector field $X=\partial_t$ is a Killing vector field for the
metric \eq{m1} on $M$, and it is well known (\emph{cf., e.g.,} 
\cite[Appendix B]{Geroch:limits}) that $X$ extends as
smoothly as the metric allows to $\pM$; we shall use the same symbol
to denote that extension. We have the following trivial
observation:
\begin{Proposition}
  \label{P34conf}
  Assume that $(\Sigma,g,V)$ is smoothly compactifiable, then
  $M=\R\times\Sigma$ with the metric \eq{m1} has a smooth conformal
  completion with $\scri$ diffeomorphic to $ \R\times\pSinfty$.
  Further $(M,\fg)$ satisfies the vacuum equations with a cosmological
  constant $\Lambda$ if and only if Equations \eq{f1}--\eq{f2} hold.
 \end{Proposition}

The implication the other way round requires some more work:
\begin{Theorem}
  \label{T4conf}
  Consider a space-time $(M,\fg)$ of the form $M=\R\times\Sigma$,
  with a metric $\fg$ of the form \eq{m1}, and suppose that there
  exists a smooth conformal completion $(\bM,\fbg)$ with nonempty
  $\scri$. Then:
\begin{enumerate}
\item $X$ is timelike on $\scri$; in particular it has no zeros there;
\item The hypersurfaces $t=$const extend smoothly to $\scri$;
\item $(\Sigma,g,V)$ is smoothly compactifiable;
\item There exists a (perhaps different) conformal completion of
  $(M,\fg)$, still denoted by $(\bM,\fbg)$, such that
  $\bM=\R\times\bS$, where $(\bS,\bg)$ is a conformal completion
  of $(\Sigma,g)$, with $X=\partial_t$ and with
  \begin{equation}
    \label{4confm}
    \fbg = -\alpha^2 dt^2 + \bg\;,
    \quad \bg(\partial_t,\cdot)=0\;,
    \quad X(\alpha)={\cal L}_X\bg = 0\;.
  \end{equation}
\end{enumerate}
\end{Theorem}
\noindent{\sc Remark:} The new completion described in point~\emph{4}.~above
    will coincide with the original one if and only if the orbits of
    $X$ are complete in the original completion.

\medskip

\noindent{\sc Proof:} As the isometry group maps $M$ to $M$, it
follows that $X$ has to be tangent to $\scri$. On $M$ we have
$\fbg(X,X)>0$ hence $\fbg(X,X)\ge0$ on $\scri$, and to establish
point~\emph{1}.~we have to exclude the possibility that $\fbg(X,X)$
vanishes 
somewhere on $\scri$.

Suppose, first, that $X(p)=0$ for a point $p\in\scri$. Clearly $X$ is
a conformal Killing vector of $\fbg$. We can choose a neighborhood
$\cU$ of $\scri$ so that $X$ is strictly timelike on
$\cU\setminus\scri$. There exists $\epsilon>0$ and a neighborhood
$\cO\subset\cU$ of $p$ such that the flow $\phi_t(q)$ of $X$ is
defined for all $q\in\cO$ and $t\in[-\epsilon,\epsilon]$. The
$\phi_t$'s are local conformal isometries, and therefore map timelike
vectors to timelike vectors. Since $X$ vanishes at $p$ the $\phi_t$'s
leave $p$ invariant. It follows that the $\phi_t$'s map causal curves
through $p$ into causal curves through $p$; therefore they map
$\partial J^+(p)$ into itself. This implies that $X$ is tangent to
$\partial J^+(p)$. However this last set is a null hypersurface, so
that every vector tangent to it is spacelike or null, which
contradicts timelikeness of $X$ on $\partial J^+(p)\cap\cU \ne
\emptyset$. It follows that $X$ has no zeros on $\scri$.

Suppose, next, that $X(p)$ is lightlike at $p$.   There exists a
neighborhood of $p$ and a 
strictly positive smooth function $\psi$ such that $X$ is a Killing
vector field for the metric $ \fbg\psi^2$.  Now the staticity
condition
\begin{equation}
  \label{statcond}
  X_{[\alpha}\nabla_\beta X_{\gamma]}=0
\end{equation}
is conformally invariant, and therefore also holds in the $\fbg $
metric. We can thus use the Carter--Vishweshvara
Lemma~\cite{CarterJMP,Vishveshwara} to conclude that the set ${\mycal N}
= \{q\in\bM|X(q)\ne 0\}\cap\partial\{ \fbg(X,X)<0\}\ne \emptyset$
is a null hypersurface.  By hypothesis there exists a neighborhood $\cU$ of
$\scri$ in $\bM$ such that ${\mycal N}\cap M\cap\cU=\emptyset$, hence
${\mycal N}\subset \scri$.  This contradicts the
fact~\cite{penrose:scri} that the conformal boundary of a vacuum
space-time with a strictly negative cosmological constant $\Lambda$
is timelike. It follows that $X$ cannot be lightlike on $\scri$
either, and point~\emph{1}.~is established.

  To establish point~\emph{2}., we note that Equation~\eq{statcond} together
  with point~\emph{1}.~show that the one-form $$\lambda\equiv{1\over{\fbg
      _{\alpha\beta}X^\alpha X^\beta}}\fbg _{\mu\nu}X^\mu dx^\nu$$ is
  a smooth closed one-form on a neighborhood $\mycal O$ of
  $\scri$, hence on any simply connected open subset of $\mycal O$ there
  exists a smooth function $\bart$ such that $\lambda=d\bart$. Now
  \eq{m1} shows that the restriction of $\lambda$ to $M$ is $dt$,
  which establishes our claim.  {}From now on we shall drop the bar on
  $\bart$, and write $t$ for the corresponding time function on
  $\bM$.

 Let
 $$\bS = \bM\cap\{t=0\}\;, \quad \chi=\Chi\Big|_{t=0}\;,\quad
 \varomega=\varOmega\Big|_{t=0}\;,
 $$ where $\Chi$ and $\varOmega$ are as in \eq{m2}; from Equation
 \eq{m2} one obtains
$$g=\chi^*(\varomega^{-2}\bg)\;,$$ which shows that $(\bS,\bg)$ is a
conformal completion of $(\Sigma,g)$. We further have $ V^2\varomega^2
= \fg(X,X)\Big|_{t=0}\varomega^2=\fbg(X,X)\Big|_{t=0} $, which has
already been shown to be smoothly extendible to $\scrip$ and strictly
positive there, which establishes point~\emph{3}.

There exists a neighborhood $\cV$ of $\bS$ in $\bM$ on which a new
conformal factor $\varOmega$ can be defined by requiring
$\varOmega\Big|_{t=0}=\varomega$, $X(\varOmega)=0$. Redefining $\fbg$
appropriately and making suitable identifications so that $\Chi$ is
the identity, Equation~\eq{m2} can then be rewritten on $\cV$ as
\begin{equation}
  \label{m3}
  \fbg = - (V\varOmega)^2 dt^2 + \varOmega^{2}g
\;.
\end{equation}
All the functions appearing in Equation~\eq{m3} are time-independent.
The new manifold $\bM$ defined as $\bS\times \R$ with the metric
\eq{m3} satisfies all the requirements of point~\emph{4}., and the proof is
complete.  \qed

In addition to the conditions described above, in
\cite{HdS,AshtekarMagnonAdS} it was proposed to further restrict the
geometries under consideration by requiring the group of conformal
isometries of $\pM$ to be the same as that of the \mK\
space-time, namely the universal covering group of $O(2,3)$;
\emph{cf.\/} also~\cite{Magnon} for further discussion.  
 While there are various ways of adapting this proposal to our setup,
   we simply note that the requirement on the group of conformal
   isometries to be $O(2,3)$ or a covering therof implies that the metric
  induced on $\pM$ is locally conformally flat.  Let
us then see what are the consequences of the requirement of local conformal
flatness of $\sg$ in our context; this last property is equivalent to
the vanishing of  
the Cotton tensor of the metric $\sg$ induced by $\fbg$ on $\pM$.  As has been discussed in detail in
Section~\ref{Sa3}, we can choose the conformal factor $\varOmega$ to
coincide with $V^{-1}$, in which case Equation~\eq{m3} reads
\begin{eqnarray}
  \fgp &\equiv &  \fg/V^{2}
\nonumber \\ & = & 
-dt^2 + V^{-2} g
\nonumber \\ & = & 
-dt^2 +  g'\;,
\label{m5}
\end{eqnarray}
with $g'\equiv V^{-2} g$ already introduced in Section~\ref{Sa3}. It
follows that
\begin{equation}
  \label{m6}
  \sg \equiv \fgp\Big|_{\scri} = -dt^2+h'\;,
\end{equation}
where $h'$ is the metric induced on $\pSinfty\equiv \pM\cap\bS$ by $g'$.
Let $\sR_{ij}$ denote the Ricci tensor of $\sg$; from \eq{m6} we obtain
\begin{equation}
  \label{m7}
  \sR_{it}=0\;, \quad \sR_{AB}={}^{2}{\cal R}_{AB}\;,
\end{equation}
where ${}^{2}{\cal R}_{AB}$ is the Ricci tensor of $h'$. In particular
the $xxA$ component of the Cotton tensor $\sR_{ijk}$ of $\sg$
satisfies
$$\sR_{xxA} = -\frac { {}^{2}{\cal R}_{,A}}4\;. $$ Point~\emph{1}.~of
Proposition~\ref{P2}, see Equation~\eq{RR}, shows that the requirement
of conformal flatness of $\sg$ implies that $R'$ is constant on
$\pSinfty$.  Conversely, it is easily seen from \eq{m7} that a locally
constant $R'$ --- or equivalently ${}^{2}{\cal R}$ --- on $\pSinfty$
implies the local conformal flatness of $\sg$. We have therefore
proved:
\begin{Proposition}
  \label{Pconf} Let $(\Sigma,g,V)$ be $C^\kind$ conformally
  compactifiable, $\kind\ge 3$, and satisfy \eq{f0}--\eq{f2}. The
  conformal boundary $\R\times\pSinfty$ of the space-time
  $(M=\R\times\Sigma,\fg)$, $\fg$ given by \eq{m1}, is locally
  conformally flat if and only if the scalar curvature $R'$ of the
  metric $V^{-2}g$ is locally constant on $\pSinfty$. This is
  equivalent to requiring that the metric induced by $V^{-2}g$ on
  $\pSinfty$ has locally constant Gauss curvature.
\end{Proposition}

\subsection{A coordinate approach}\label{Sa4c}

An alternative approach to the conformal one discussed above is by
introducing preferred coordinate systems. As discussed in
\cite[Appendix~D]{HT}, coordinate approaches are often equivalent to
conformal approaches when sufficiently strong hypotheses are made.  We
stress that this equivalence is a delicate issue when finite degrees
of differentiability are assumed, as arguments leading from one
approach to the other often involve constructions in which some
differentiability is lost.

In any case, the coordinate approach has been used by Boucher, Gibbons
and Horowitz~\cite{BGH} in their argument for uniqueness of the
\mK\ metric within a certain class of static space-times.
More precisely, in~\cite{BGH} one considers metrics which are
asymptotic to \gK\ metrics with $k=1$ in the following strong sense:
if $g_0$ denotes one of the metrics \eq{Kot} with $k=1$, then one assumes
that there exists a coordinate system $(t,r,x^A)$ such that 
\begin{eqnarray}
  g&=&g_0 + O(r^{-2}) dt^2 + O(r^{-6}) dr^2 \nonumber \\ & & \; +
  O(r)\,\mbox{(remaining differentials not involving $dr$)} \nonumber
  \\ & & \; + O(r^{-1})\,\mbox{(remaining differentials involving
    $dr$)}\;.
  \label{bgh}
\end{eqnarray}
We note that in the uniqueness assertions of~\cite{BGH} one makes
appeal to the positive energy theorem to conclude. Now we are not
aware of a version of such a theorem which would hold without some
further hypotheses on the behavior of the metric. 
For example, in such a theorem one is likely to require that the
derivatives of the 
metric also fall off at some sufficiently high rates. In any case the
argument presented in~\cite{BGH} seems to implicitly assume that the
asymptotic behavior of $g^{tt}$ described above is preserved under
differentiation, so that the corrections terms in \eq{bgh} give a
vanishing contribution when calculating $|dV|^2_g -|dV_0|^2_{g_0}$ and
passing to the limit $r\to\infty$, with $g_0$ --- the \mK\
metric. While it might well be possible that Equations
\eq{f1}--\eq{f2} force the metrics satisfying \eq{bgh} to have
sufficiently good asymptotic properties to be able to justify this, or
to apply a positive energy theorem\footnote{Recall that in the
  asymptotically flat case one can derive an asymptotic expansion for
  stationary metrics from rather weak hypotheses on the leading order
  behavior of the
  metric~\cite{KennefickMurchadha,Chnohair,SimonBeig}.  See
  especially~\cite{manderson:stationary,manderson:static}, where the
  Lichnerowicz theorem is proved without any hypotheses on the
  asymptotic behavior of the metric, under the condition of geodesic
  completeness of space-time.}, this remains to be
established.\footnote{The key point of the
  argument in \cite{BGH} is to prove that the coordinate mass is
  negative. When $\pSinfty=S^2$, and the asymptotic conditions are
  such that the positive energy theorem applies, one can conclude that
  the initial data set under consideration must be coming from one in
  anti-de Sitter space-times \emph{provided} one shows that the
  coordinate mass coincides with the mass which occurs in the positive
  energy theorem. To our knowledge such an equality has not been
  proved so far for metrics with the asymptotics \eq{bgh}, or else.}

It is far from being clear whether or not a general metric of the form
\eq{bgh} has any well behaved conformal completions. For example, the
coordinate transformation~\eq{defr} together with a multiplication
by the square of the conformal factor $\varomega= x$ brings
the metric \eq{bgh} to one which can be continuously extended to the
boundary, but if only \eq{bgh} is assumed then the resulting metric
will not be differentiable up to boundary on the compactified manifold
in general. There could, however, exist coordinate systems which lead
to better conformal behavior when Equations \eq{f1}--\eq{f2}
are imposed.

In any case, it is natural to ask whether or not a metric satisfying
the requirements of Section~\ref{Sa3} will have a coordinate
representation similar to \eq{bgh}. A partial answer to this question
is given by the following\footnote{See~\cite[Appendix]{HT} for a
  related discussion.  While the conclusions in~\cite{HT} appear to be
  weaker than ours, it should be stressed that in~\cite{HT} staticity
  of the space-times under consideration is not assumed.}:

\begin{Proposition}
  \label{Pcoord} Let $(\Sigma,g,V)$ be a $C^\kind$ compactifiable
  solution of Equations \eq{f1}--\eq{f2}, $\kind\ge 3$. Define a $C^{i-2}$
  function $\newk=\newk(x^A)$ on $\pSinfty$ by the formula
  \begin{equation}
    \label{again}
    R'|_{\pSinfty}= -2\Lambda  \newk\;.
  \end{equation}
  \begin{enumerate}
  \item Rescaling $V$ by a positive constant if necessary, there
    exists a coordinate system $(r,x^A)$ near $\pSinfty$ in which we
    have
  \begin{eqnarray}
    \label{gcoord}
&   V^2= \frac{r^2}{\ell^2}+\newk\;,&\\
&  g = \displaystyle \frac{dr^2}{\left(\frac{r^2}{\ell^2} + \newk -
    \frac{2\mu}{r}\right)}  +O(r^{-3})dr\, dx^A
+ \left({r^2}\ch_{AB} 
+O(r^{-1})\right) dx^Adx^B
& \label{gcoord1}
 \end{eqnarray}
 (recall that $\ell^2 = - 3 \Lambda^{-1}$), for some $r$-independent smooth
 two-dimensional metric $\ch_{AB}$ with Gauss curvature equal to $\newk$
  and for some  function $\mu=\mu(r,x^A)$. 
  Further
 \begin{eqnarray}
 & \ch^{AB} g_{AB}= 2\left(r^2-\displaystyle\frac
   {\mu_{\infty}}r+O(r^{-2})\right)\;, 
\label{traceg}\end{eqnarray}where $ \ch^{AB}$ denotes the matrix
inverse to $ \ch_{AB}$ while 
\begin{eqnarray} 
& \mu_{\infty}\equiv\lim_{r\to\infty}\mu= \displaystyle\frac{\ell^3}{12}
   \frac{\partial R'}{\partial 
     x}\Big|_{x=0}\;.&
  \label{gcoord2} \end{eqnarray}
\item \label{Pcoordp2} If one moreover assumes  that $R'$ is locally
 constant on $\pSinfty$, then Equation~\eq{gcoord1} can be improved to
 \begin{eqnarray}
 &  g =  \displaystyle{\frac{dr^2}{\left(\frac{r^2}{\ell^2} + k -
    \frac{2\mu}{r}\right)}}
+ \left({r^2} \ch_{AB}
+O(r^{-1})\right) dx^Adx^B\;,& \label{gcoord3}
 \end{eqnarray}
 with $\ch_{AB}$ having constant Gauss curvature $k=0,\pm1$ according
 to the genus of the connected component of $\pSinfty$ under
 consideration.
  \end{enumerate}

\end{Proposition}
\noindent {\sc Remarks:} 1. The function $(x,x^A)\to\mu(r=1/x,x^A)$
is of differentiability class $C^{i-3}$ on $\bS$, with the function
$(x,x^A)\to(\mu/r)(r=1/x,x^A)$ being of differentiability class $C^{i-2}$ on
$\bS$. 

2. In Equations~\eq{gcoord1} and \eq{gcoord3}
the error terms 
  $O(r^{-j})$ satisfy
  $$\partial_r^s\partial_{A_1}\ldots
  \partial_{A_t}O(r^{-j})=O(r^{-j-s})$$
  for $0\le s+t \le i-3$.

3. We emphasize that the function $\newk$ defined in
Equation~\eq{again} could \emph{a priori} be $x^A$-dependent. In such
a case neither the definition of coordinate mass of
Section~\ref{Scoord} nor the definition of Hamiltonian mass of
Section~\ref{Shamil} apply.

4. It seems that to be able to obtain \eq{bgh}, in addition to the
hypothesis that $R'$ is locally constant on $\pSinfty$ one would at
least need the quantity appearing at the right hand side of
Equation~\eq{gcoord2} to be locally constant on $\pSinfty$ as well. We
do not know whether this is true in general; we have not investigated
this question as this is
irrelevant for our purposes.

\medskip

\proof
Consider, near $\pSinfty$, the coordinate system of Equation~\eq{gprime};
from Equations \eq{pprime} and \eq{Rprime5} we obtain
\begin{equation}
  \label{xlog}
  \partial_x\left(\ln \sqrt{\det h'_{AB}}\right) = -2\newk x -
\frac{3\mu_\infty}\ell x^2 +O(x^3)\;,
\end{equation}
$\ell$ as in \eq{Lshort}, $\newk$ as in \eq{again}, $\mu_\infty$ as in
\eq{gcoord2}.  This, together with Equation~\eq{qfall3}, leads to
\begin{eqnarray*}
  \lefteqn{\frac{\partial h'_{AB}}{\partial x} = -2x\newk h'_{AB}+
  O(x^2)\quad \Longrightarrow  }
\\
& & \phantom{xxxxxxxxxxxx} {h'_{AB}= (1-\newk x^2)\ell^2\ch_{AB} + O(x^3)}\;,
\end{eqnarray*}
where
$$\ch_{AB}\equiv \frac1{\ell^2}h'_{AB}\Big|_{x=0}\;.$$
Proposition~\ref{P2} shows
that $\newk$ is proportional to the Gauss curvature of $\ch_{AB}$. It
follows now from \eq{Rprime5} that 
$$g=x^{-2}g'=
\frac{\ell^2}{x^2\left(1-\frac{R'\ell^2x^2}{6}\right)}dx^2 + \left\{
  \frac{(1-\newk x^2)}{x^2}h'_{AB}\Big|_{x=0} +O(x^3)
\right\}dx^Adx^B\;. $$
The above suggests to introduce a coordinate
$r$ via the formula\footnote{We note that $\newk$ is of
  differentiability class lower by two orders as compared to the
  metric itself, which leads to a loss of three derivatives when
  passing to a new coordinate system in which $r$ is defined by
  Equation~\eq{rdef}. One can actually introduce a coordinate system closely
  related to \eq{rdef} with a loss of only one degree of
  differentiability of the metric by using the techniques of
  \cite[Appendix A]{AndChDiss}, but we shall not discuss this here.}
\begin{equation}
  \label{rdef}
  \frac{r^2}{\ell^2} = \frac{1-\newk x^2}{x^2}\;.
\end{equation}
Suppose, first, that $\newk $ is locally constant on $\pSinfty$, then $\newk$ equals $k=0,\pm 1$ according to the genus of the
connected component of $\pSinfty$ under consideration, and one finds
\begin{eqnarray*}
  g & = & \frac{dr^2}{\left(\frac{r^2}{\ell^2} + k\right)\left\{
1 + \frac{\ell^2}{r^2}\left(k-\frac{R'\ell^2x^2}{6} \right)
\right\}} + \left(\frac{r^2}{\ell^2} h'_{AB}\Big|_{x=0}
+O(r^{-1})\right) dx^Adx^B
\\ & = & \frac{dr^2}{\left(\frac{r^2}{\ell^2} + k -
    \frac{2\mu}{r}\right)} 
+ \left(\frac{r^2}{\ell^2} h'_{AB}\Big|_{x=0}
+O(r^{-1})\right) dx^Adx^B\;,
\end{eqnarray*}
where the ``mass aspect'' function $\mu=\mu(r,x^A)$ is defined as
\begin{eqnarray}
  \mu& \equiv& -\frac r2 \left(1+k \frac{\ell^2}{r^2}\right)
  \left(k-\frac{R'\ell^2x^2}{6} \right)\nonumber \\ & = & -\frac r2
  \left(k -\frac{R'\ell^2}{6} + \frac{k^2\ell^2}{r^2}\right)
  \nonumber\\ & = & \frac {r\ell^2}2 \left(\frac{1}{6}(R'-R'|_{x=0})
    - \frac{k^2}{r^2}\right)\;.
\label{mass(r)}
\end{eqnarray}
This establishes Equations \eq{gcoord} and
\eq{gcoord3}. When $\newk$ is not locally
constant an identical  calculation using the coordinate $r$ defined in
Equation~\eq{rdef} establishes Equation~\eq{gcoord1} --- the only
difference is the occurrence of non-vanishing error terms in the $dr
dx^A$ part of the metric, introduced by the angle dependence of $\newk$. 
It follows from  Equation~\eq{mass(r)} --- or from the $\newk$ version thereof 
when $\newk$ is not locally constant --- 
that 
\begin{eqnarray*}
  \mu & = & \frac{\ell^3}{12} \frac{\partial R'}{\partial x}\Big|_{x=0}
  + O(r^{-1})\;, 
\end{eqnarray*}
which establishes Equation~\eq{gcoord2}.  
Equation~\eq{traceg} is obtained by integration of Equation~\eq{xlog}.

\section{Connectedness of $\pSinfty$}
\label{Sconn}

The class of manifolds considered so far could in principle contain
$\Sigma$'s for which neither $\pSinfty$ nor $\pSo$ are connected.
Under the hypothesis of staticity the question of connectedness of
$\pSo$ is open; we simply note here the existence of dynamical
(non-stationary) solutions of Einstein--Maxwell equations with a
non-connected black hole region with positive cosmological constant
$\Lambda$ \cite{BHKJ,KastorTraschen}. As far as $\pSinfty$ is concerned, 
we have the following:

\begin{Theorem}
  \label{Tconn} Let $(\Sigma,g,V)$ be a $C^\kind$ compactifiable
  solution of Equations \eq{f1}--\eq{f2}, $\kind\ge 3$. Then
  $\pSinfty$ is connected.
\end{Theorem}

\proof 
  Consider the manifold
$M=\R\times\Sigma$ with the metric \eq{m1}; its conformal completion
$\bM=\R\times\bS$ with the metric $\fg/V^{2}$ is a stably causal
manifold with boundary. We wish to show that it is also globally
hyperbolic in the sense of \cite{GSWW}, namely that 1) it is strongly
causal and 2) for each $p,q\in M$ the set $J^+(p)\cap J^-(q)$ is
compact. The existence of the global time function $t$ clearly implies
strong causality, so it remains to verify the compactness condition.
Now a path $\Gamma(s)=(t(s),\gamma(s))\in \R\times\Sigma$ is an
achronal null geodesic from $p=(t(0),\gamma(0))$ to
$q=(t(1),\gamma(1))$ if and only if $\gamma(s)$ is a minimizing
geodesic between $\gamma(0)$ and $\gamma(1)$ for the ``optical
metric'' $V^{-2} g$. Compactness of $J^+(p)\cap J^-(q)$ is then
equivalent to compactness of the $V^{-2} g$-distance balls; this
latter property will hold when $(\Sigma\cup\pSinfty,V^{-2} g)$ is a
geodesically complete manifold (with boundary) by (an appropriate
version of) the Hopf--Rinow theorem.

Let us thus show that $(\Sigma,V^{-2} g)$ is geodesically complete.
Suppose, first, that $\partial\Sigma=\emptyset$; the hypothesis that
$\Sigma$ has compact interior together with the fact that $V$ tends to
infinity in the asymptotic regions implies that $V\ge V_0>0$ for some
constant $V_0$. This shows that $(\Sigma,V^{-2} g)$ is a compact
manifold with boundary $\pSinfty$, and the result follows. (When the metric
induced by $V^{-2} g$ on $\pSinfty$ has positive scalar curvature
connectedness of $\pSinfty$ can also be inferred from
\cite{WittenYau}.)

Consider, next, the case $\pSo\ne\emptyset$. It is well known that
$|dV|_g$ is a non-zero constant on every connected component of
$\pSo$ (\emph{cf.} the discussion around Equation \eq{Wvanish});
therefore we can introduce coordinates near $\pSo$ so that $V=x$, with
the metric taking the form
\begin{equation}
  \label{boundme}
  V^{-2} g = x^{-2} \left( (dx)^2 + h_{AB}(x,x^A) dx^A dx^B\right)\;,
\end{equation}
where the $x^A$'s are local coordinates on $\pSo$. It is elementary to
show now from \eq{boundme} that $(\Sigma\cup\pSinfty,V^{-2} g)$ is
a complete manifold with boundary, as claimed.

When $(\Sigma,g)$ is smoothly compactifiable we can now use
\cite[Theorem~2.1]{GSWW} to infer connectedness of $\pSinfty$, compare
\cite[Corollary, Section~III]{GSWW2}. For compactifications with
finite differentiability we argue as follows:  For small $s$ let 
$\lambda$ be the mean curvature of the sets $\equiv\{x=s\}$, where $x$ 
is the coordinate of Equation~\eq{Gausscoord}. In  
the coordinate system used there the unit normal to those sets
pointing away from $\pSinfty$ equals
$x\partial_x$; if $(\Sigma,g,V)$ is $C^3$ compactifiable the tensor field
$\bh$ appearing in Equation~\eq{Gausscoord} will be $C^1$ so
that\footnote{The differentiability threshold $k=3$ can be lowered using
  the ``almost Gaussian coordinate systems'' of \cite[Appendix
  A]{AndChDiss}, we shall however not be concerned with this here.}
\begin{eqnarray*}
  \lambda & = & \frac 1{\sqrt{\det g}}\partial_i\left(\sqrt{\det g}\;
    n^i\right)
\\ &=& \frac {x^3}{\sqrt{\det \bh}} \partial_x\left(x^{-2}\sqrt{\det \bh}
    \right)
\\ &=& -2 + O(x)\;.
\end{eqnarray*}
It follows  that for $s$ small enough the sets
$\{x=s,t=\tau\}$ are trapped, with respect to the
inward pointing normal, in the space-time $\R\times\Sigma$ with the
metric \eq{m1}. 
Suppose that $\pSinfty$ were not connected, then those (compact) sets would
be outer trapped with respect to every other connected component of
$\pSinfty$. This is, however, not possible by the usual global
arguments, \emph{cf., e.g.}, \cite{Gannon1,Gannon2} or
\cite[Section~4]{ChDGH} for details. \qed

\section{The mass}\label{Smass}
\subsection{A coordinate mass $M_c$}
\label{Scoord}

There exist several proposals how to assign a mass $M$ to a
space-time which is asymptotic to an anti-de Sitter space-time
\cite{HT,AshtekarMagnonAdS,AbbottDeser,GibbonsGPI,AshtekarDas}; it
seems that the simplest way to do that (as well as to extend the definition
to the \gK\ context considered here) proceeds as follows: consider a
metric defined on a coordinate patch covering the set
\begin{equation}
  \label{Sext}
  \Sext\equiv\{t=t_0,r\ge R,(x^A)\in {}^2M\}
\end{equation}
(which we will call an ``end''), and suppose that in this coordinate
system the functions $g_{\mu\nu}$ are of the form \eq{Kot} plus lower
order terms\footnote{Because the
  natural length of the vectors $\partial_A$ is $O(r)$ it would actually
  be natural to require $g_{r\mu}=o(r)$, $\mu\ne r,t$ instead
  of $g_{r\mu}=o(1)$, $\mu\ne r,t$.} 
\begin{eqnarray}
 & g_{tt}=-(k - \frac {2m}r - \frac \Lambda 3
r^2)+\frac {o(1)}{r}\;,\qquad g_{rr}=1/(k - \frac {2m}r - \frac \Lambda
3 r^2+\frac {o(1)}{r})\;,& \nonumber\\ &
 g_{t\mu}=o(1)\;,\quad \mu\ne t\;,\qquad g_{r\mu}=o(1)\;,\quad \mu\ne r,t\;,
& \nonumber\\ &  g_{AB}-r^2h_{AB} = o(r^2)\;,\label{eq:coordmasscond}
\end{eqnarray}
for some constant $m$, and for some  constant
curvature ($t$ and $r$ independent) metric
$h_{AB}dx^Adx^B$ on ${}^2M$.  Then we
define the coordinate mass $M_c$ of  
the end $\Sext$ to be the parameter $m$ appearing in \eq{Kot}.  This
procedure gives a unique prescription how to assign a mass to a metric
\emph{and a coordinate system} on $\Sext$.

There is an obvious coordinate-dependence in this definition when
$k=0$: Indeed, in that case a coordinate transformation $r\to\lambda
r$, $t\to t/\lambda$, $d\Omegak^2\to \lambda^{-2} d\Omegak^2$, where
$\lambda$ is a positive constant, does not change the asymptotic form
of the metric, while $m$ gets replaced by $\lambda^{-3} m$.  When
${}^2M$ is compact this freedom can be removed 
\emph{e.g.} by requiring
that the area of ${}^2M$ with respect to the metric $d\Omegak^2$ be
equal to $4\pi$, or to $1$, or to some other chosen constant. For
$k=\pm 1$ this ambiguity does not arise because in this case such
rescalings will change the asymptotic form of the metric, and are
therefore not allowed.

It is far from being clear that the above definition will assign the
same parameter $M_c$ to every coordinate system satisfying our
requirements: 
if that is the case, to prove such a statement one might
perhaps need to further require that the coordinate derivatives of the
coordinate components of $g$ in the above described coordinate system
have some appropriate decay properties; further one might perhaps have
to replace the $o(1)$'s by $o(r^{-\sigma})$'s or $O(r^{-\sigma})$'s,
for some appropriate $\sigma$'s, perhaps as in \eq{bgh}; this is
however irrelevant for our discussion at this stage.

A possible justification of this definition proceeds as follows: when
$\Mtwo= S^2$ and $\Lambda=0$ it is widely accepted that the mass of
$\Sext$ equals $m$, because $m$ corresponds to the active
gravitational mass of the gravitational field in a quasi-Newtonian
limit. (It is also known in this case that the definition is
coordinate-independent \cite{ChErice,bartnik:mass}.)  For $\Lambda\ne 0$
and/or $\Mtwo\ne S^2$ one calls $m$ the mass by extrapolation.

Consider, then, the metric \eq{m1}, with $V$ and $g$ as in
\eq{gcoord}--\eq{gcoord1}; suppose further that the limit
$$\mu_\infty\equiv \lim_{r\to\infty} \mu$$ exists and is a constant. 
 To achieve the form
of the metric $\fg$ just described one needs to replace the coordinate
$r$ of \eq{gcoord}--\eq{gcoord1} with a new coordinate $\rho$ defined
as
$$r^2+k=\rho^2+k+\frac{\mu_\infty}\rho\;.$$
This leads to 
  \begin{eqnarray}
    \fg& = & -\left(\frac{\rho^2}{\ell^2} + k +
      \frac{\mu_\infty}{\rho}\right)dt^2+
    \frac{d\rho^2}{\left(\frac{\rho^2}{\ell^2} + k +
        \frac{\mu_\infty}{\rho}+O(\frac1{ \rho^{2}})\right)} \nonumber\\ & &
    \label{mas1} \qquad +O(\rho^{-3})d\rho\, dx^A + \left({\rho^2} \ch_{AB}
      +O(\rho^{-1})\right) dx^Adx^B\;,
 \end{eqnarray}
and therefore 
\begin{equation}
  \label{coordmass}
  M_c\equiv -\frac {\mu_\infty} 2 = -\frac{\ell^3}{24} \frac{\partial
    R'}{\partial x}\Big|_{x=0}\;,
\end{equation}
where the second equality above follows from \eq{gcoord2}. We note
that the approach described does not give a definition of mass when
$\lim_{r\to\infty} \mu$ does not exist, or is not a constant function
on $\pSinfty$.

The above described dogmatic approach suffers from various
shortcomings. First, when $\Mtwo\ne S^2$, the arguments given are
compatible with $M_c$ being any function $M_c(m,\Lambda)$ with the
property that $M_c(m,0)=m$. Next, the transition from $\Lambda\ne 0$
to $\Lambda =0$ has dramatic consequences as far as global properties
of the corresponding space-times are concerned, and one can argue
that there is no reason why the function $M_c(m,\Lambda)$ should be
continuous at zero. For example, according to
\cite[Equation~(III.8c)]{HT}, the mass of the metric \eq{Kot} with
$\Mtwo=S^2$ should be $16\pi m \ell$, with $\ell$ as in \eq{Lshort},
which blows up when $\Lambda$ tends to zero with $m$ being held
fixed. Finally, when 
$\Mtwo\ne S^2$ the Newtonian limit argument does not apply because the
metrics \eq{Kot} with $\Lambda=0$ and $\Mtwo\ne S^2$ do not seem to
have a Newtonian equivalent. In particular there is no reason why
$M_c$ should not depend upon the genus $g_\infty$ of $\Mtwo$ as well.

All the above arguments make it clear that a more fundamental approach
to the definition of mass would be useful. It is common  in
field theory to define energy by Hamiltonian methods, and this is the
approach we shall pursue in the next section.

\subsection{The Hamiltonian mass $\mham$.}
\label{Shamil}
The Hamiltonian approach allows one to define the energy from first
principles. For a solution of the field equations, we can simply
take as the energy the numerical value of the Hamiltonian.
It must be recognized, however, that the Hamiltonians 
might depend on the choice of the phase space, if several such 
choices are possible, and they are defined only up to an additive
constant on each connected component of the phase space. 
They also depend on the choice of the Hamiltonian structure, 
if more than one such structure exists.

Let us start by briefly recalling the results of the analysis of
\cite{ChAIHP}, based on the Hamiltonian approach of Kijowski and
Tulczyjew \cite{KijowskiTulczyjew,KijowskiGRG}, see also
\cite{Kijowskiold}.  One assumes that a manifold $M$ on which an
(unphysical) background metric $\backg$ is given, and one
considers metrics $\fg$ which asymptote to $\backg$ in the relevant
asymptotic regions of $M$.  We stress that the background metric is
 only a tool to prescribe the asymptotic boundary conditions, and does 
not have any  physical significance. 
Let $X$ be any vector field on
$M$ and let $\Sigma$ be any hypersurface in $M$.  By a well known
procedure the motion of $\Sigma$ along the flow of $X$ can be used to
construct a Hamiltonian dynamical system in an appropriate phase space
of fields over $\Sigma$; the reader is referred to
\cite{KijowskiTulczyjew,KijowskiGRG,Kijowskiold,CJK} for a geometric
approach to this question. In \cite{ChAIHP} it is also assumed that
$X$ is a Killing vector field of the background metric; this is
certainly not necessary (\emph{cf., e.g.}, \cite{CJK} for general
formulae),
but is sufficient for our purposes, as we are going to take $X$ to be
equal to $\partial/\partial t$ in the coordinate system of Equation
\eq{m1}. In the context of metrics which asymptote to the \gK\ 
metrics at large $r$, a rigorous functional description of the spaces
involved has not been carried out so far, and lies outside the scope
of this paper. Let us simply note that one expects, based on the
results in \cite{Friedrich:adS,Kannar:adS,CJK}, to obtain a well
defined Hamiltonian system in this context.  Therefore the formal
calculations of \cite{ChAIHP} lead one to expect that on an
appropriate space of fields, such that the associated physical
space-time metrics $\fg$ asymptote to the background metric $\backg$
at a suitable rate, the Hamiltonian $H(X,\Sigma)$ will 
coincide with (or, at worse, will be closely related to) the one given by the
formula derived in \cite{ChAIHP}:
\begin{eqnarray}
  H(X,\Sigma )&= &\frac 12 \int_{\partial\hyp}
 \ourU^{\alpha\beta}dS_{\alpha\beta}\;, 
\label{toto}
\end{eqnarray}
where the integral over $\partial \hyp$ should be understood by a
limiting process, as the limit as $R$ tends to infinity of integrals
of coordinate spheres $t=0$, $r=R$ on $\Sext$. Here $d
S_{\alpha\beta}$ is defined as
$\frac{\partial}{\partial x^\alpha}\lrcorner \frac{\partial}{\partial
  x^\beta}\lrcorner \rd x^0 \wedge\cdots \wedge\rd x^{n} $, with
$\lrcorner$ denoting contraction, and $\ourU^{\alpha\beta}$ is given by

\begin{eqnarray}
  \ourU^{\nu\lambda}&= &
{\ourU^{\nu\lambda}}_{\beta}X^\beta + \frac 1{8\pi} \left(\sqrt{|\det
g_{\rho\sigma}|}~g^{\alpha[\nu}-\sqrt{|\det b_{\rho\sigma}|}~
b^{\alpha[\nu}\right)\delta^{\lambda]}_\beta {X^\beta}_{;\alpha}
\ ,\label{Fsup2new}
\\ {\ourU^{\nu\lambda}}_\beta &= & \displaystyle{\frac{2|\det
  \bmetric_{\mu\nu}|}{ 16\pi\sqrt{|\det g_{\rho\sigma}|}}}
g_{\beta\gamma}(e^2 g^{\gamma[\nu}g^{\lambda]\kappa})_{;\kappa}
\;.\label{Freud2.0} 
\end{eqnarray}
Here, and \emph{throughout this section}, $g$ stands for the space-time
metric $\fg$ unless explicitly indicated otherwise. Further, a
semicolon  denotes covariant 
differentiation \emph{with respect to the background metric $b$},
while
\begin{equation}
  \label{mas2}
e\equiv \frac{\sqrt{|\det
g_{\rho\sigma}|}}{\sqrt{|\det\bmetric_{\mu\nu}|}}\; .
\end{equation}
Some comments concerning Equation~\eq{Fsup2new} are in order: in
\cite{ChAIHP} the starting point of the analysis is the Hilbert
Lagrangian for vacuum Einstein gravity, 
$$\mathcal{L}= \sqrt{- \det g_{\mu\nu}}~\frac{g^{\alpha \beta}
  R_{\alpha \beta}}{16\pi}\;. $$  As the normalization factors 
play an important role in giving a correct definition of mass, we recall
that  the
factor $1/16\pi$ is  determined by the requirement that the theory
has the correct Newtonian limit (units $G=c=1$ are used
throughout). With our signature $(-+++)$ the 
Einstein equations with a cosmological constant read
$$R_{\mu\nu}-\frac{g^{\alpha \beta} R_{\alpha \beta}}{2} g_{\mu\nu} =
-\Lambda g_{\mu\nu} \;, $$
so that one needs to repeat the analysis in \cite{ChAIHP} with
$\mathcal{L}$ replaced by
$$\frac{\sqrt{- \det g_{\mu\nu}}}{16\pi}\left(g^{\alpha \beta}
  R_{\alpha \beta} -2\Lambda\right)\;.$$ 
The general expression for the Hamiltonian \eq{toto} in terms of
$X^{\mu}$, $g_{\mu\nu}$ and $b_{\mu\nu}$  ends up to 
coincide with that obtained with $\Lambda=0$,
which can be seen either by direct calculations, or by the Legendre
transformation arguments of \cite[end of Section~3]{ChAIHP} together
with the results in \cite{Kijowskiold}.  Note that Equation~\eq{Fsup2new}
does not exactly coincide with that derived in \cite{ChAIHP}, as the
formulae there do not contain the term $-\sqrt{|\det
  b_{\rho\sigma}|}~b^{\alpha[\nu}\delta^{\lambda]}_\beta
{X^\beta}_{;\alpha}$. However, this term does not depend on the
metric $g$, and such terms can be freely added to the Hamiltonian
because they do not affect the variational formula that defines a
Hamiltonian. {}From an energy point of view such an addition corresponds
to a choice of the zero point of the energy.  We note that in our
context $H(X,\Sigma )$ would not converge if the term $-\sqrt{|\det
  b_{\rho\sigma}|}~b^{\alpha[\nu}\delta^{\lambda]}_\beta
{X^\beta}_{;\alpha}$ were not present in \eq{Fsup2new}.

In order to apply this formalism in our context, we let $\backg$ be
any $t$-independent metric on $M=\R\times\Sigma$ such that
(with $0 \ne \Lambda = - 3/\ell^2$) 
\begin{equation}
  \label{bKot}
  \backg = 
 -(k + \frac{r^2}{\ell^2}) dt^2 + (k + \frac{r^2}{\ell^2})^{-1} dr^2 
+ r^2 \ch 
\end{equation}
on $\R\times\Sext\approx \R\times[R,\infty)\times {}^2M$, for some $R\ge 0$, 
where $\ch=\ch_{AB}dx^Adx^B$ denotes a metric of constant Gauss curvature
$k=0,\pm 1$ on the two dimensional connected compact manifold ${}^2M$.

Let us return to the discussion in
Section~\ref{Scoord} concerning the freedom of rescaling the
coordinate $r$ by a positive constant $\lambda$. First, if $k$ in
Equation~\eq{bKot} is any constant different from zero, then there
exists a (unique) rescaling of $r$ and $t$ which brings $k$ to one, or
to minus one. Next, if $k= 0$ we can --- without changing the asymptotic
form of the metric --- rescale the coordinate $r$ by a positive
constant $\lambda$, the coordinate $t$ by $1/\lambda$, and the metric
$\ch$ by $\lambda^{-2}$, so that there is still some freedom left in
the coordinate system above; a unique normalization can then be
achieved by asking \emph{e.g.} that the area
\begin{equation}
  \label{areanorm}
  A_\infty\equiv \int_{\Mtwo} d^2\mu_{\ch}
\end{equation}
equals $4\pi$ --- this will be the most convenient normalization for
our purposes. Here $d^2\mu_{\ch}$ is the Riemannian measure associated
with the metric $\ch$. We wish to point out that that \emph{regardless} of
the value of $k$, the Hamiltonian $H(X,\Sigma)$ is \emph{independent}
of this scaling: this follows immediately from the fact that
$\ourU^{\alpha\beta}$ behaves as a density under linear coordinate
transformations. An alternative way of
seeing this is that 
in the new coordinate system $X$ equals $\lambda \partial/\partial t$,
which accounts for a factor $1/\lambda$ in the transformation law of
the coordinate mass, as discussed at the beginning of Section
\ref{Scoord}. The remaining factor $1/\lambda^2$ needed there is
accounted for by a change of the area of $\pSinfty$ under the
rescaling of the metric $\ch$ which accompanies that of $r$.

In order to evaluate $H$ it is useful to introduce the following
$\backg$-orthonormal frame:
\begin{equation}
  \label{frame}
  e_{\zero}= \frac{1}{\sqrt{k + \frac{r^2}{\ell^2}}}\partial_t\;,
  \quad e_{\one}=  {\sqrt{k + \frac{r^2}{\ell^2}}} \partial_r\;, \quad
  e_{\A} = \frac 1r \ce_{\A}\;,
\end{equation}
where $\ce_\A$ is an ON frame for the metric $\ch$. 
The connection coefficients, defined by the formula
$\nabla_{e_{\ha}}e_{\hb}={\omega^{\hatc}}_{\hb\ha}e_{\hatc}$, read
\begin{eqnarray}
  & \omega_{\zero \one \zero} = - \frac r {\ell^2 \sqrt{k + \frac
      {r^2}{\ell^2}}} = -\frac 1 \ell + O(r^{-2})\;, & 
\nonumber \\ & \omega_{
    \one \two \two} = \omega_{ \one \three \three} = - \frac 
  {\sqrt{k + \frac{r^2}{\ell^2}}} r = -\frac 1 \ell + O(r^{-2})\;, &
\nonumber \\ 
  & \omega_{ \two \three \three} = \cases{ -\frac{\coth \theta}r \;, &
    $k=-1\;,$ \cr 0 \;, & $k=0\;,$ \cr -\frac{\cot \theta}r \;, &
    $k=1\;.$} & \label{connection}
\end{eqnarray}
Those connection coefficients which are not obtained from the above ones
by permutations of indices are zero; we have used a coordinate system
$\theta,\varphi$ on $\Mtwo$ in which $\ch$ takes, locally, the form
$d\theta^2 + \sinh^2 \theta\; d\varphi^2$ for $k=-1$, $d\theta^2 +
d\varphi^2$ for $k=0$, and $d\theta^2 + \sin^2 \theta \;d\varphi^2$ for
$k=1$.  We also have
 \begin{eqnarray}
   \label{Killing}
& X^{\zero}= \sqrt{k + \frac{r^2}{\ell^2}} = \frac r {\ell} +
O(r^{-1})\;,& \\ \label{Killing1}  
& e_{\one}(X^{\zero})={X^{\zero}}_{;\one} = - X_{\zero;\one} = X_{\one;\zero} 
= \frac r {\ell^2} 
\;,&
 \end{eqnarray}
 with the third equality in \eq{Killing1} following from the Killing
 equations $X_{\mu;\nu}+ X_{\nu;\mu}=0$; all the remaining
 $X^{\hmu}$'s and $X_{\hmu;\hnu}$'s are zero.  Let the tensor field
 $e^{\mu\nu}$ be defined by the formula
\begin{equation}
  \label{mas3}
  e^{\mu\nu}\equiv g^{\mu\nu}-b^{\mu\nu}
\; .
\end{equation}
We shall use hatted indices to denote the components of a tensor field 
in the frame $e_{\ha}$ defined in \eq{frame}, \emph{e.g.} $e^{\ha\hc}$
denotes the coefficients of $e^{\mu\nu}$ with respect to that
frame:
$$e^{\mu\nu}\partial_\mu\otimes \partial_\nu=e^{\ha\hc}e_{\ha}\otimes
e_{\hc}\;.$$
Suppose that the metric $\fg$ is such that the
$e^{\ha\hc}$'s tend to zero as $r$ tends to infinity. By a
Gram--Schmidt  procedure we can find a frame $\gf_{\ta}$,
$\ta=0,\ldots,3$, orthonormal \emph{with respect to the metric $g$},
such that $\gf_{0}$ is proportional to $e_{0}$, and such
that the $e_{\ha}$ components of $\gf_{0}-e_{0}$, $\ldots$,
$\gf_{3}-e_{3}$ tend to zero as $r$ tends to infinity:
\begin{equation}
  \label{eq:framedecay}
  \gf_{\ta}={\gf_{\ta}}^{\ha}e_{\ha} 
\to_{r\to\infty}\delta_{\ta}^{\ha}e_{\ha}\;.  
\end{equation}
{}From \eq{toto} and \eq{eq:framedecay} we expect that\footnote{
  Equation \eq{toto1} will indeed turn out to be correct under the
  conditions \eq{hamfalof} imposed below.}
\begin{eqnarray}
  H(X,\Sigma )&= &\lim_{R\to\infty}  \int_{\Sigma\cap\{r=R\}}
 r^2\ourU^{\one\zero} d^2\mu_{r}\;, 
\label{toto1}
\end{eqnarray}
where $d^2\mu_{r}$ is the Riemannian measure induced on
${\Sigma\cap\{r=R\}}$ by $\fg$. We wish to analyze when the above
limit exists; we have
$$r^2{\ourU^{\one\zero}}_{\beta}X^\beta =
r^2{\ourU^{\one\zero}}_{\zero}X^{\zero}\approx
\frac{r^3}\ell{\ourU^{\one\zero}}_{\zero}\;, $$ hence we need to keep
track of all the terms in ${\ourU^{\one\zero}}_{\zero}$ which decay 
as $r^{-3}$ or slower. Similarly one sees from Equations
\eq{Killing}--\eq{Killing1} that only those terms in
$$\Delta^{\halpha\hnu}\equiv\sqrt{|\det g_{\hrho\hsigma}|}~
g^{\halpha\hnu}-\sqrt{|\det b_{\hrho\hsigma}|}~b^{\halpha\hnu}$$ which
are $O(r^{-3})$, or which are decaying slower, will give a
non-vanishing contribution to the term involving the derivatives of
$X$ in the integral \eq{toto1}. This suggests to consider metrics
$\fg$ such that
\begin{equation}
  \label{hamfalof}
  e^{\hmu\hnu}= o(r^{-3/2})\;, \quad e_{\hrho}(e^{\hmu\hnu})= o(r^{-3/2})\;.
\end{equation}
The boundary conditions \eq{hamfalof} ensure that one needs to keep
track only of those terms in $\ourU^{\one\zero}$ which are linear in
$e^{\hmu\hnu}$ and $e_{\hrho}(e^{\hmu\hnu})$, when $\ourU^{\one\zero}$ is
Taylor expanded around $\backg$. For a \gK\ metric \eq{Kot} we have
    \begin{equation}
      \label{egk}
      e^{\zero\zero}\approx e^{\one\one} \approx
      -\frac{2m\ell^2}{r^3}\;,\qquad e_{\one}(e^{\zero\zero}) \approx
      e_{\one}(e^{\one\one}) \approx \frac{6m\ell}{r^3}\;,
    \end{equation}
    with the remaining $e^{\hmu\hnu}$'s and
    $e_{\hsigma}(e^{\hmu\hnu})$'s vanishing, so that Equations
    \eq{hamfalof} are   satisfied. Under \eq{hamfalof} one
    obtains
    \begin{eqnarray}
g_{\ha\hc} & = & \eta_{\ha\hc}-\eta_{\ha\hr}\eta_{\hc\hs}
e^{\hr\hs}+o(r^{-3})\;,
\label{metricdown} \\
\sqrt{|\det g_{\mu\nu}|} & = & \sqrt{|\det \backg_{\mu\nu}|}\left(1 +
  \frac 12 (e^{\zero\zero} -e^{\one\one}-
    e^{\hA\hA})+o(r^{-3})\right)\;,
\nonumber \\
      {\ourU^{\one\zero}}_{\zero} & = & -\frac1{16\pi}
      \left(2e_{;\one}+ {e^{\one\hi}}_{;\hi}-
        {e^{\zero\zero}}_{;\one}\right) +o(r^{-3}) \nonumber\\ & = &
      \frac1{16\pi} \left(e_{\one}( e^{\hA\hA})+ \frac 1 \ell
        (e^{\hA\hA}-2e^{\one\one} ) - \frac1 r{\check\mathcal{ D}}_{\hA}
        e^{\one\hA}\right) +o(r^{-3}) \;,
     \nonumber
      \\ \frac1{8\pi} \Delta^{\alpha[\one}
      {X^{\zero]}}_{;\alpha} & = & \frac1{16\pi}
      \left(\Delta^{\one\one}-\Delta^{\zero\zero}\right){X^{\zero}}_{;\one}
      \nonumber\\ & = & \frac r
      {16\pi\ell^2}\left(\Delta^{\one\one}-\Delta^{\zero\zero}\right)
      +o(r^{-3})\nonumber\\ & = &  -\frac r {16\pi\ell^2}e^{\hA\hA} +o(r^{-3})
  \label{uo1oX} \;.   \end{eqnarray}
The indices $\hi$ run from $1$ to $3$ while the indices $\hA$ run from
$2$ to $3$; ${\check\mathcal{ D}}_{\hA}$ denotes the covariant
derivative on $\Mtwo$, and ${\check\mathcal{ D}}_{\hA} e^{\one\hA}$ is
understood to be the covariant derivative associated with the metric
$\ch$ of a vector field on $\Mtwo$, 
with repeated $\hA$ indices being summed over.  In Equation
\eq{metricdown} $\eta_{\hmu\hnu}=$diag$(-1,+1,+1,+1)$, while the
$g_{\hmu\hnu}$'s are the components of the tensor $g_{\hmu\hnu}$ in a
co-frame dual to \eq{frame}. Inserting all this into \eq{toto1} one
is finally led to the  simple expression
 \begin{eqnarray}
   \mham&\equiv& H(\frac{\partial}{\partial t}, \{t=0\}) \nonumber \\ 
   &= & \lim_{R\to\infty} \frac {r^3}
   {16\pi\ell^2}\int_{\Sigma\cap\{r=R\}} \left(r{\frac {
         \partial e^{\hA\hA}}{\partial r}}-2e^{\one\one}
   \right) d^2\mu_{\ch}\;.
   \label{massequation}
 \end{eqnarray}
 In particular if $\fg$ is the \gK\ metric \eq{Kot} one obtains
 (\emph{cf.\/} Equation~\eq{egk})
\begin{equation}
  \label{masskot}
  \mham= \frac{A_\infty m}{4\pi}\;,
\end{equation}
$A_\infty$ defined in \eq{areanorm}. If 
$\Mtwo=T^2$ with area normalized to $4\pi$ we obtain $\mham=m$. For
$k=\pm 1$ it follows from the Gauss--Bonnet theorem that
$A_\infty=4\pi|1-g_\infty|$, where $g_\infty$ is the genus of $\Mtwo$,
hence
\begin{equation}
  \label{masskot2}
  \mham= |1-g_\infty| m\;.
\end{equation}
 This gives again $\mham=m$ for $\Mtwo=S^2$,
but this will not be true anymore for $\Mtwo$'s of higher
genus. We believe that the Hamiltonian approach is the one which provides
\emph{the} correct definition of mass in field theories, and therefore
Equations \eq{masskot}--\eq{masskot2} are the ones which provide the
correct normalization of mass. 

Let us finally consider static metrics $\fg$ of the form \eq{m1}, and
suppose that the hypotheses of point~\emph{\ref{Pcoordp2}}.~of
Proposition~\ref{Pcoord} hold. We can then use the coordinates of that
proposition to calculate $\mham$, and obtain
\begin{equation}
  \label{eq:stathammas}
\mham=-\frac{1}{8\pi}\int_{\pSinfty}{\mu_\infty}\; d^2\mu_{\ch}\;.  
\end{equation}
If we further assume that $\mu_\infty$ is constant on $\pSinfty$,
Equation~\eq{eq:stathammas} gives
$$\mham=-\frac{\mu_\infty} 2=M_c$$
for $\Mtwo=S^2$ and for an appropriately
normalized $T^2$, while $$
\mham=
-|1-g_\infty| \frac{\mu_\infty} 2=|1-g_\infty| M_c$$
for higher genus $\pSinfty$'s. Here $M_c$ is the coordinate mass as
defined in Section~\ref{Scoord}.

\subsection{A generalized Komar mass}
\label{SKomar}
Recall that the Komar mass is a number which can be assigned to every
stationary, asymptotically flat metric the energy-momentum tensor of
which decays sufficiently rapidly:
\begin{equation}
  \label{eq:Km}
  M_K=\lim_{R\to\infty} \frac 1 {8\pi} \int_{S_{R,T}} \sqrt{|\det
    g_{\alpha \beta}|}\; \nabla^\mu 
  X^\nu \;dS_{\mu\nu}\;,
\end{equation}
where $X^\mu\partial_\mu$ is the Killing vector field which asymptotes
to $\partial/\partial t$ in the asymptotically flat region, and the
$S_{R,T}\equiv\{t=T,r=R\}$'s are coordinate spheres in that region.
The normalization factor $1/(8\pi)$ has been chosen so that $M_K$
reproduces the familiar mass parameter $m$ when Schwarzschild metrics
are considered. For metrics considered here with $\Lambda\ne 0$ the
integral \eq{eq:Km} diverges when $X^\mu\partial_\mu=\partial/\partial
t$ and when the $S_{R,T}$'s are taken to be coordinate spheres in the
region $\Sext$ where the metric exhibits the \gK\ asymptotics. An
obvious way to generalize $M_K$ to the situation considered in this
paper is to remove the divergent part of the integral using a
background metric $\backg$:
\begin{equation}
  \label{eq:Km1}
  M_K=\lim_{R\to\infty} \frac 1 {8\pi} \int_{S_{R,T}}
  \left(\sqrt{|\det g_{\alpha \beta}|}\; \nabla^\mu 
  X^\nu - \sqrt{|\det \backg_{\alpha \beta}|}\; \bnabla ^\mu X^\nu
  \right)dS_{\mu\nu}\;. 
\end{equation}
Here $\bnabla$ denotes a covariant derivative with respect to the
background metric. More precisely, let $\Sext$, $\backg$,
$\ch$, \emph{etc}., be as in Equation~\eq{bKot}, and consider
time-independent metrics 
$g$ which in the coordinate system of Equation~\eq{bKot} are of the
form \eq{m1} with
\begin{eqnarray}
&V^2=\frac{r^2}{\ell^2}+ \newk  -\frac{2\beta}r +o(\frac 1r)\;, 
& \nonumber \\ &
\partial_r(V^2-\frac{r^2}{\ell^2}- \newk  +\frac{2\beta}r)=o(\frac 1{r^2}) 
\;, & \nonumber 
\\ & g^{rr}=\frac  {r^2}{\ell^2}+ \newk  -\frac{2\gamma}r + o(\frac 1
{r} )\;, & \nonumber \\ &\sqrt{|\det g_{\alpha \beta}|} =
\left( r^2+\frac{2\delta\ell^2}r+o(\frac 1r)\right)\sqrt{|\det \ch_{AB}|}\;, 
&\label{eq:Km2}
\end{eqnarray}
for some $r$-independent differentiable functions $\newk=
\newk(x^A)$, $\beta=\beta(x^A)$, $\gamma=\gamma(x^A)$ and
$\delta=\delta(x^A)$ defined on a coordinate neigbhourhood of
$\pSinfty$.
(The conditions \eq{eq:Km2} roughly reflect the behavior 
of the metric in the coordinate system of Proposition~\ref{Pcoord}).
Under \eq{eq:Km2} the limit as $R$ tends to infinity in the definition
\eq{eq:Km1} of $M_K$ exists, and one finds
\begin{eqnarray}
  M_K&=&\lim_{R\to\infty}\frac 1 {4\pi}\int_{S_{R,T}}\left(\sqrt{|\det
      g_{\alpha \beta}|}\; g^{r\mu} g^{\nu t}\partial_{[\mu}g_{\nu] t} - 
\sqrt{|\det \backg_{\alpha \beta}|}\; \backg^{r\mu} \backg^{\nu
  t}\partial_{[\mu}\backg_{\nu] t}  
  \right)dx^2 dx^3 \nonumber 
\\&=&\lim_{R\to\infty}\frac 1 {8\pi}\int_{S_{R,T}}\left(\sqrt{|\det g_{\alpha \beta}|}\; g^{rr} g^{tt}\partial_{r}g_{t t} - 
\sqrt{|\det \backg_{\alpha \beta}|}\; \backg^{rr} \backg^{tt}\partial_{r}\backg_{t t} 
  \right)dx^2 dx^3 \nonumber 
\\&=& \frac 1 {4\pi}\int_{\pSinfty} (3\beta-2\gamma+2\delta)\; d^2\mu_{\ch}\;.
  \label{eq:km3}
\end{eqnarray}
It turns out that the value of $M_K$ so obtained depends on the
background metric chosen. (Our definition of background,
Equation~\eq{bKot}, is tied to the choice of a particular coordinate
system, so another way of stating this is that the number $M_K$ as
defined so far is assigned to a metric \emph{and} to a coordinate
system, in a manner somewhat similar to the coordinate mass of
Section~\ref{Scoord}). Indeed, given any differentiable function
$\alpha(x^A)$ there exists a neighborhood of $\pSinfty$ on which a new
coordinate $\hr$ can be introduced by the formula
\begin{eqnarray}
  \label{eq:Km4}
  \frac{\hr^2}{\ell^2} - 2\frac\alpha \hr = \frac{r^2}{\ell^2}
\;. 
\end{eqnarray}
We can then chose the new background to be $ \backg = -(k + \frac{\hr^2}
{\ell^2}) dt^2 + (k + \frac{\hr^2}{\ell^2})^{-1} d\hr^2 + \hr^2
\ch$, and obtain a new $M_K$ which will in general \emph{not} coincide
with the old one. (It is noteworthy that the coordinate transformation
\eq{eq:Km4} with the associated change of background do \emph{not}
change the value of the Hamiltonian mass  
$\mham$.) For example, if $\alpha$ is constant, using
hats to denote the corresponding functions appearing in the metric
expressed in the new coordinate system we obtain
  \begin{eqnarray*}
&    \hat\beta=\beta + \alpha\;, \qquad \hat\gamma = \gamma+3\alpha\;,
    \qquad \hat\delta=\delta -2 \alpha & \\
& \Longrightarrow \qquad\hat M_K = M_K- \frac {7\alpha A_\infty} {4\pi}\;, 
  \end{eqnarray*}
  where $A_\infty$ is the area of $\pSinfty$ with respect to the
  metric $\ch$.  It turns out that one can remove this coordinate
  dependence in an appropriate class of metrics, tailoring the
  prescription in such a way that Equation~\eq{eq:km3} reproduces, up
  to a genus dependent factor, the coordinate mass $M_c$. In order to
  do that we shall suppose that the metric $\fourg$ satisfies the
  hypotheses of point~\emph{\ref{Pcoordp2}}.~of Proposition~\ref{Pcoord} (in
  particular $\newk=k=0,\pm1$ according to the genus of the connected
  component of $\pSinfty$ under consideration), and we let the
  background be associated with a coordinate system $(\rho,x^A)$ with
  $\rho$ given by \eq{gcoord}. It follows from Equations~\eq{mas1}
  and \eq{traceg} that in this coordinate system it holds
  \begin{equation}
    \label{eq:km5}
    \sqrt{|\det
    g_{\alpha \beta}|}\; = r^2 +o(\frac 1 r)\;,
  \end{equation}
where we have used the generic symbol $r$ to denote the coordinate $\rho$.
We then {\em impose \eq{eq:km5} as a restriction on the coordinate
  system in which the generalized Komar mass $M_K$ has to be
  calculated.} When this condition is imposed we obtain from \eq{mas1} 
and \eq{eq:stathammas}
$$M_K=-\frac{1}{8\pi}\int_{\pSinfty}{\mu_\infty}\;
d^2\mu_{\ch}=\mham\;. $$
We have thus proved
\begin{Proposition}
  \label{PKomar}
  Consider a metric $\fourg$ satisfying the hypotheses of
  point~\emph{\ref{Pcoordp2}}.~of Proposition~\ref{Pcoord}, then the
  generalized Komar mass \eq{eq:Km1} associated to a background metric
  \eq{bKot} such that \eq{eq:km5} holds equals the Hamiltonian mass
  \eq{massequation}.
\end{Proposition}

Proposition \ref{PKomar} is the $\Lambda<0$ analogue of the theorem of
Beig~\cite{BeigKomar}, that for static $\Lambda=0$ vacuum metrics
which are asymptotically flat in spacelike directions the ADM mass and
the Komar masses coincide. Our treatment here is inspired by, and
somewhat related to, the analysis of \cite{Magnon}.

\subsection{The \GHn\ mass $\mghp$}
\label{Sghm}

Let $\psi$ be a function defined on the asymptotic region $\Sext$,
with $\Sext$ defined as in \eq{Sext}, such that the level sets of
$\psi$ are smooth compact surfaces diffeomorphic to each other (at
least for $\psi$ large enough), with $\psi\to_{r\to\infty}\infty$.
Following Hawking \cite{SWH}, Gibbons \cite[Equation~(17)]{GibbonsGPI}  
assigns a mass $\mghp$ to such a foliation via the formula
\begin{equation}\label{dHmn}
\mghp 
\equiv  \mbox{lim}_{
  \epsilon \to 0} \frac{\sqrt{A_{1/\epsilon}}}{32 \pi^{3/2}}
\int_{\{\psi=1/\epsilon\}}( {}^{2}{\cal R} - \frac{1}{2}p^{2}
-\frac{2}{3}\Lambda) dA \; ,
\end{equation}
where $ A_{\alpha}$ is the area of the connected component under
consideration of the level set $\{\psi=\alpha\}$

By considering simple examples in Minkowski space-times it can be
seen that this definition is $\psi$ dependent. However, when
$\Mtwo=S^2$, $\Lambda=0$, and the coordinate system on $\Sext$ is such
that the ADM mass $m_{ADM}$ (which equals $m_H$ as defined in Section
\ref{Shamil}) of $\Sext$ is well defined (see
\cite{bartnik:mass,ChErice}), then $\mghp$ will be independent of
$\psi$, in the class of $\psi$'s singled out by the condition that the
level sets of $\psi$ approach round spheres at a suitable
rate. No results of this
kind are known when $\Lambda\ne 0$.

The definition \eq{dHmn} applied to the
function $\psi=r$ and the metric \eq{Kot} with $k\ne 0$ gives
$$\mgh= 
{m}{|1-g_\infty|^{3/2}}\;.$$ We have also used the Gauss--Bonnet
theorem to calculate $\sqrt{A_{1/\epsilon}}$. Thus the definition
\eq{dHmn} differs from the coordinate one by the somewhat unnatural
factor $|1-g_\infty|^{3/2}$.  It is not clear why such a factor should
be included in the definition of mass.

Consider, next,
the metrics \eq{m1} with $V$ and $g$ given by
\eq{gcoord}--\eq{gcoord1}. Let $\psi=V$; from the Codazzi--Mainardi
Equation~\eq{CMnn}, the Equation~\eq{f2}, and the definition
\eq{Wdef} of $W$ we obtain, for $V$ large enough so that $|dV|>0$,
  \begin{eqnarray*}
    {}^2\cR - \frac 12 p^2 - \frac 23 \Lambda & = & (-2 R_{ij} +R
    g_{ij}) n^i n^j - |q_{ij}|^2_g - \frac 23 \Lambda
\\ & = & -2 \frac{D^i VD^j V}{VW} D_i D_j V - |q_{ij}|^2_g - \frac 23 \Lambda
\\ & = & - \frac{D^i VD_i W}{VW} - |q_{ij}|^2_g - \frac 23 \Lambda\;.
  \end{eqnarray*}
In the coordinate system of Equation~\eq{gprime}, where $V=1/x$, one
is led to
  \begin{eqnarray*}
    {}^2\cR - \frac 12 p^2 - \frac 23 \Lambda & = & x^3 
    \frac{\partial W}{\partial x}  - \frac 23 \Lambda +O(x^6)
\\ & = &  -\frac{x^3}6    \frac{\partial R'}{\partial x}  +O(x^6)\;,
  \end{eqnarray*}
  and we have used \eq{qfall2} and \eq{Rprime}. {}From
  $A_{1/\epsilon}\approx x^{-2}A^\prime _{\pSinfty}$ we finally
  obtain
\begin{eqnarray}
   \mgh(V) & =  & -\frac{\sqrt{A^\prime_{\pSinfty}}}{32 \pi^{3/2}}\int_{\pSinfty} 
\frac 16\frac{\partial R'}{\partial x} d^2\mu_{h^\prime}
\nonumber \\ & = &
 -\frac{\sqrt{A^\prime_{\pSinfty}}}{32 \pi^{3/2}}\int_{\pSinfty} 
\ell\frac{ n'(R')}{6} d^2\mu_{h^\prime}
\;,
 \label{mghgk}
\end{eqnarray}
where $d^2\mu_{h^\prime}$ is the Riemannian area element induced by
$g'$ on $\pSinfty$, and $n'$ denotes the inward-pointing $g'$-unit
normal to $\pSinfty$.
We have thus proved the following result:
\begin{Theorem}
  \label{Tmgh} Let a triple $(\Sigma,g,V)$ satisfying Equations
  \eq{f0}--\eq{f2} be $C^\kind$ compactifiable, $\kind \ge 3$. Then
  the \GHn\ mass $\mgh(V)$ of the $V$-foliation is finite and well
  defined; it is given by the formula \eq{mghgk}, with $R'$ --- the
  curvature scalar of the metric $g'=V^{-2}g$.
\end{Theorem}
 We can relate  $\mgh(V)$ to the coordinate mass $M_c$ if we assume
in addition that the latter is well defined; recall that this required
$R'$ and $\partial _x R'$ to be constant on $\pSinfty$. In this case
Equation~\eq{coordmass} gives
\begin{equation}
  \label{massrel}
  \mgh(V)= \left(\frac{{A'_{\pSinfty}}}{4 \pi\ell^2}\right)^{3/2}M_c\;.
\end{equation}
{}From Equation~\eq{RR} we have $  {}^{2}{\cal R'}|_{x=0}=
{2k}/{\ell^2}$, and the Gauss--Bonnet theorem implies
$$\int_{\pSinfty} {}^{2}{\cal R'}
d^2\mu_{h^\prime}=\frac{2k}{\ell^2}{A'_{\pSinfty}} =
8\pi(1-g_\infty)\;, $$ so that when $g_\infty\ne 1$ we obtain
\begin{equation}
  \label{massrelghc}
  \mgh(V)= |1-g_\infty|^{3/2} M_c\;.
\end{equation}
We emphasize that $\mgh(V)$ is finite and well defined even when the
conditions of Section \eq{Scoord}, which we have set forth to define
$M_c$, are not met.

Similarly, 
the Hamiltonian mass $\mham$, associated to the background singled out
by the coordinate system of Proposition~\ref{Pcoord}, can be
defined when $R'$ is constant on $\pSinfty$. (This holds regardless of
whether or not  $\partial _x R'$ is 
constant on $\pSinfty$.) Proceeding as above, making use of
Equations~\eq{again}--\eq{gcoord3}, one is led to 
\begin{eqnarray}
\nonumber &
 g_\infty\ne 1 \quad\Longrightarrow\quad \mgh(V)= |1-g_\infty|^{1/2} \mham\;,
 & \\ &
 g_\infty= 1\;, \ A'_\infty = 4\pi\ell^2  \quad\Longrightarrow\quad
 \mgh(V)=  \mham\;. 
 \label{massrelghc2}\end{eqnarray}

\section{The generalized Penrose inequality}
\label{GPI}
We recall here an argument of  Geroch~\cite{Geroch:extraction}, as
extended by Jang and Wald \cite{JangWald} and Gibbons \cite{GibbonsGPI}, 
for the validity of the Penrose inequality\footnote{The  argument we
  review has been used by Gibbons in \cite{GibbonsGPI} to
 obtain a somewhat different inequality, in which the genus factors
 are not present. The inequality in \cite{GibbonsGPI} is violated for
 \gK\ metrics with $g_\infty\ge 3$.}: 
\begin{Proposition}
\label{PHIP}
Assume we are given a three dimensional manifold ${(\Sigma,g)}$ with
 connected boundary $\partial \Sigma$ such that: 
\begin{enumerate}
\item $R \ge 2\Theta$ for some strictly negative constant $\Theta$.
\item There exists a smooth, global solution of the inverse mean
  curvature flow  without critical points,  
\emph{i.e.}, there exists a surjective function
$u:\Sigma\to[0,\infty)$ such that $du$ has no zeros and
\begin{equation}
  \label{uf}
\cases{u|_{\pSo}=0\;,\cr
\nabla_i\left(\frac{\nabla^i u }{|du|}\right)= |du| \;.}  
\end{equation}
\item The level sets of $u$
$$N_s = \{u(x)=s\}\;$$
are compact. 
\item
 The boundary $\partial \Sigma = u^{-1}(0)$ of $\Sigma$ is minimal.
\item
The \GHn\ mass  of the level sets of $u$ as defined
in \eq{dHmn} exists.  
\end{enumerate}
Then
\begin{equation}
\label{rds1x}
2  \mgh(u) \ge 
  (1-g_{\pSo}) \left(\frac{A_{\pSo}}{4\pi}\right)^{1/2}- \frac
  \Theta 3  \left(\frac{A_{\pSo}}{4\pi}\right)^{3/2} \;.
 \end{equation}Here  $A_{\pSo }$
is the area of $\pSo $ and $g_{\pSo }$ is the genus thereof.
\end{Proposition}

\noindent{\sc Remarks:}
1. The Proposition above can be applied to solutions of \eq{f1}
and \eq{f2} with $\Theta=\Lambda$:
 in this case we have $R = 2 \Lambda$; further Equation~\eq{f2}
 multiplied by $V$ and  contracted with two vectors tangent to $\pSo $
shows that the boundary $\{V = 0\}$ is totally geodesic and hence
minimal. 

2. Equation~\eq{rds1x} is sharp --- the inequality there becomes an
 equality for the \gK\ metrics.

\medskip\proof
Let $A_s$ denote the area of $N_s$, and define
\begin{equation}
  \label{sigdef}
  \sigma(s) = \sqrt{A_s} \int_{N_s} ({}^2\cR_s - \frac12 p_s^2 - \frac 23
  \Theta) d^2\mu_s\;,
\end{equation}
where ${}^2\cR_s$ is the scalar curvature of the metric induced on
$N_s$, $d^2\mu_s$ is the Riemannian volume element associated to that
same metric, and $p_s$ is the mean curvature of $N_s$. 
The hypothesis that $du$ is nowhere vanishing implies that all the
objects involved are smooth in $s$. At $s=0$ we have
\begin{eqnarray}
  \sigma(0)&=& 
\sqrt{A_{\pSo}} \int_{\pSo} ({}^2\cR_0 -
  \frac 23 \Theta) d^2\mu_0 
\nonumber \\ & = & \sqrt{A_{\pSo}}
  \left(8\pi(1-g_{\pSo}) - \frac 23 \Theta A_{\pSo}\right) \;.
\label{firstGPI}\end{eqnarray}
On the other hand,
$$\lim_{s\to\infty}\sigma(s)= 32 \pi^{3/2}\mghu\;.$$ 
The  generalization in \cite{GibbonsGPI} of the
classical calculation of \cite{Geroch:extraction} 
gives
\begin{equation}
  \label{monoton}
\frac{  \partial \sigma}{\partial s }\ge 0\;.
\end{equation}
This  implies
$\lim_{s\to\infty}\sigma(s)\ge \sigma(0)$, which gives \eq{rds1x}.
\qed

 To be able to carry out the above argument one had to assume that
 $du$ had no zeros, which implies in particular that $\pSinfty$ is
 connected with $g_{\pSo}=g_{\infty}$. It is not known whether or not
 the  other hypotheses of Proposition~\ref{PHIP}, or the conditions of
 Definition~\ref{Dc} together with Equations~\eq{f0}--\eq{f2}, force
 $\pSo$ to be connected. If they do not, 
 one is tempted to conjecture that the right inequality should be 
\begin{equation}
\label{rdso}
2  \mgh(u) \ge \sum_{i=1}^k
  \left((1-g_{\pSi}) \left(\frac{A_{\pSi}}{4\pi}\right)^{1/2}- \frac
  \Theta 3  \left(\frac{A_{\pSi}}{4\pi}\right)^{3/2}\right) \;.
 \end{equation}
Here the $\pSi$'s, $i=1,\ldots,k$, are the connected
components of $\pSo$, $A_{\pSi }$
is the area of $\pSi $, and $g_{\pSi }$ is the genus thereof.
  This would be the inequality one would obtain from the
  Geroch--Gibbons argument if it could be carried through for $u$'s
  which are allowed to have critical points, on manifolds with
  $\pSinfty$ connected but $\pSo$  --- not connected.

 We note that when $\Lambda=0$ there is a
 version of the proof of Proposition~\ref{PHIP} due to Huisken and
 Ilmanen in which $du$ is allowed 
 to have zeros (with $\pSo$ --- connected)\footnote{Bray's
   proof  \cite{Bray:preparation2} of the inequality~\eq{rdso} with
   $\Theta=0$ but $\pSo$ --- not necessarily connected, uses a completely 
   different technique; in particular it makes appeal to the positive
   energy theorem which does not hold in the class of manifolds
   considered here.}. Note that at points where $du$ vanishes
 Equation~\eq{uf} does not make sense classically, and has to be
 understood in a proper way. Further the monotonicity calculation of
 \cite{Geroch:extraction} breaks down at critical level sets of $u$,
 as those do not have to be smooth submanifolds.  Nevertheless (when
 $\Lambda=0$) existence of appropriate functions $u$ (perhaps with
 critical points) together with the monotonicity of $\sigma$ can be
 established \cite{HI1,HI2} when $\pSo$ is an outermost (necessarily
 connected) minimal sphere. It is conceivable 
 that the argument of Huisken and Ilmanen can be modified to include
 the case $\Lambda\ne 0$. One of the difficulties here is to handle
 the possibly changing genus of the level sets of $u$.
 
  Let us discuss some of the consequences of the (hypothetical) Equation \eq{rdso}. To
 proceed further it is convenient to introduce a mass parameter $m$
 defined as follows:
\begin{equation}
  \label{masspar}
  m = \cases{ \mgh\;, & $\pSinfty=S^2\;,$ \cr \mgh\;, &
    $\pSinfty=T^2$, with the normalization $A'_\infty = 4\pi\ell^2 \;,$ \cr
    \displaystyle\frac{\mgh}{|g_{\pSinfty} -1|^{3/2}}\;, &
    $g_{\pSinfty} > 1 \;.$ }
\end{equation}
Strictly speaking, we should write $m(u)$ if $\mgh(u)$ is used above,
$m(V)$ if $\mgh(V)$ is used, \emph{etc.}; we shall do this when
confusions are likely to occur. {}For \gK\ metrics the mass $m=m(u)$
so defined coincides with the mass parameter appearing in \eq{Kot}
when $u$ is the radial solution $u=u(r)$ of the problem \eq{uf};
$m(V)$ coincides with the coordinate mass $M_c$ for the metrics
considered here when $M_c$ is defined, \emph{cf.\/}
Equation~\eq{massrel}.

Note, first, that if all connected components of the horizon have
spherical or toroidal topology,  
then the lower bound \eq{rdso} is strictly positive.
For example, if  $\pSo=T^2$, and  $\pSinfty=T^2$ as well we obtain
$$2m\ge \frac1{\ell^2}\left(\frac{A_{\pSo}}{4\pi}\right)^{3/2}\;.$$
On the other hand if $\pSo=T^2$ but $g_{\pSinfty} >1$  from
Equation~\eq{rdso} one obtains 
$$2m\ge \frac1{\ell^2|g_\infty-1|} \left(\frac{A_{\pSo}}
  {4\pi}\right)^{3/2}\;.$$   
Let us return to the case\footnote{The discussion that follows actually applies to all
  $(\Sigma,g)$'s that can be isometrically embedded into a 
globally hyperbolic space-time $M$ in which the null convergence
condition holds; further the image of $\Sigma$ should be a partial
Cauchy surface in  $M$. Finally the intersection of $\Sigma$ with
$\Scri$ should be compact. The global hyperbolicity here, and the notion
of Cauchy surfaces, is understood in 
the sense of manifolds with boundary, see \cite{GSWW} for details.}  where
Equations~\eq{f0}--\eq{f2} hold; we can then use the
Galloway--Schleich--Witt--Woolgar inequality
\cite{GSWW}
\begin{equation}
  \label{gsww}
  \sum_{i=1}^k{g_{\pSi}}\le g_\infty\;.
\end{equation}
It implies that if $\pSinfty$ has spherical topology, then all connected
components of the horizon must be spheres. Similarly, if $\pSinfty$ is a
torus, then all components of the horizon are spheres, except perhaps
for at most one which could be a torus. It follows that to have a
component of the horizon which has genus higher than one we need
$g_\infty >1$ as well.

When some --- or all --- connected components of the horizon have genus
higher than one the right hand side of Equation~\eq{rdso} might become
negative. Minimizing the generalized Penrose inequality \eq{rdso} with
respect to the areas of the horizons gives the following interesting
inequality
\begin{equation}
  \label{iPi1}
 \mgh(u) \ge -\frac{1}{3 \sqrt{-\Lambda}} \sum_i |g_{\pSi} -1|^{3/2}\;,
\end{equation}
where the sum is over those connected components $\pSi$ of $\pSo$ for
which $g_{\pSi} \ge1$.  Equation~\eq{iPi1}, 
 together with the elementary
inequality $\sum_{i=1} ^N |\lambda_i|^{3/2} \le \left(\sum_{i=1} ^N
  |\lambda_i|\right)^{3/2}$, lead to
\begin{equation}
  \label{iPi3}
  m \ge -\frac{1}{3 \sqrt{-\Lambda}} \;.
\end{equation}

The Geroch--Gibbons argument establishing the inequality \eq{firstGPI}
when a suitable $u$ exists can also be \emph{formally} carried through
when $\pSo=\emptyset$. In this case one still considers solutions $u$
of the differential equation that appears in Equation~\eq{uf}, however
the Dirichlet condition on $u$ at $\pSo$ is replaced by a condition on
the behavior of $u$ near some chosen point $p_0\in\Sigma$. If the
level set of $u$ around $p_0$ approach distance spheres centered at
$p_0$ at a suitable rate, then $\sigma(s)$ tends to zero when the
$N_s$'s shrink to $p_0$, which together with the monotonicity of
$\sigma$ leads to the positive energy inequality:
\begin{equation}
  \label{gpet}
  \mgh(u)\ge 0\;.
\end{equation}
It should be emphasized that the Horowitz-Myers solutions
\cite{HorowitzMyers} with negative mass show
that this argument breaks down  when $g_\infty=1$.

When $\pSinfty=S^2$ one expects that \eq{gpet}, with $\mgh(u)$ replaced
by the spinorially defined mass (which might perhaps coincide with
$\mgh(u)$, but this remains to be established), can be proved by
Witten type techniques, compare \cite{AndDahl,GHHP}. On the other hand
it follows from \cite{Baum} that when $\pSinfty\ne S^2$ there exist no
asymptotically covariantly constant spinors which can be used in the
Witten argument. The Geroch--Gibbons argument has a lot of ``ifs''
attached in this context, in particular if $\pSinfty\ne S^2$ then some
level sets of $u$ are necessarily critical and it is not clear what
happens with $\sigma$ when a jump of topology from a sphere to a
higher genus surface occurs. We note that the area of the horizons
does not occur in \eq{iPi3} which, when $g_{\pSinfty}>1$, 
suggests that the correct inequality is actually \eq{iPi3} rather than
\eq{gpet}.

\section{Mass and area inequalities}
\label{spT1}

\subsection{Preliminaries}

We first give here a sketch of the content of this section.
We define $W_0$ via a suitably chosen generalized Kottler solution, 
(called the "reference solution", RS)  and 
$\widetilde W = \Psi^{-4} W$ and $\widetilde W_0 = \Psi^{-4} W_0$ for a
certain function $\Psi(V)$. We then establish three lemmas. The first one 
(Lemma \ref{LHm}) expresses the surface integral at infinity 
of the normal derivative  $n^{i}D_{i}(\widetilde W - \widetilde W_{0})$ 
in terms of the mass difference between the given solution and a RS,
 while Lemma \ref{Lahor} expresses this same normal derivative taken on the 
horizon in terms of the difference of the areas of the given and the RS,
with appropriate genus factors. 
We next recall from ~\cite{BeigSimon3}, an elliptic equation of the form
$(\triangle - a)(\widetilde W - \widetilde W_{0}) \ge 0$, for some function
$a$. This equation is first employed in Lemma \ref{Lrigidity} where we show that the 
generalized Kottler solutions can be characterized either by the condition
 $\widetilde W = \widetilde W_{0}$ or by conformal flatness of $(\Sigma, g)$
(this Lemma is actually formulated more genrerally such as to include the
Nariai case as well). 
The crucial step in the proofs then consists of applying the maximum principle 
to  the elliptic equation for $\widetilde W - \widetilde W_{0}$. This is
possible if the function $a$ is non-negative, which is the case in the present 
situation ($\Lambda < 0$) iff the mass of the reference Kottler solution is 
non-positive. By the asymptotic conditions (and by a suitable choice of the
RS) we can achieve that $\widetilde W - \widetilde W_{0}$ takes its maximum 
value (namely zero) both at the horizon (if there is one) and at infinity. 
The maximum principle then yields that the  derivatives 
$ n^{i}D_{i}(\widetilde W - \widetilde W_{0})$ 
 with respect to the outward normals at the horizon and at infinity are 
positive, and zero precisely if $\widetilde W = \widetilde
W_{0}$. Theorems \ref{T1.1}  and \ref{T1.2} then readily follow from the 
lemmata.  As a  final step we combine the mass and area inequalities to derive 
the inverse Penrose inequality.

Turning, then,  to the details, let $(\Sigma,g,V)$ satisfy \eq{f0}--\eq{f2} 
together with the topological, the differential, and the asymptotic 
requirements spelled out in the statements of Theorems~\ref{T1.1}
or \ref{T1.2}. (As mentioned above, Lemma \ref{Lrigidity} holds under more
general conditions). We first introduce the surface gravity $\kappa$ 
of $\pSo$ to be the corresponding restriction of the function $\sqrt{W}$ 
defined by~\eq{Wdef} to
$\pSo $:
\begin{equation}
  \label{surfacegravity}
  \kappa \equiv |dV|_g\Big|_{\pSo }\;.
\end{equation}
where we have normalized $V$ so that Equation~\eq{Rprime0} holds, 
\emph{cf.\/} Proposition~\ref{P2}. By the strong maximum 
principle~\cite[Lemma~3.4]{GT} $W$ is nowhere vanishing on $\pSo$. 
Moreover, it is well known (and easily seen using Equation~\eq{f2}) 
that $\kappa$ is locally constant on $\pSo $:
\begin{eqnarray}\nonumber
  0 = n^jD_iD_j V \Big|_{V=0}& = & \frac{D^jV}{\sqrt{W}}D_iD_j
  V\Big|_{V=0}
\\ & = & \frac{1}{2\sqrt{W}}D_iW \Big|_{V=0}\ . \label{Wvanish}
\end{eqnarray}
Here $n^i$ is the unit normal to $\pSo $, where $V$ vanishes.

It is convenient to introduce the notion of a {\it reference solution}
(RS) as follows: 
This is a generalized Kottler solution with the same genus $g_{\infty}$
as $(\Sigma,g,V)$. Moreover, if $\pSo \ne \emptyset$, the surface gravity 
$\kappa$ of the RS is chosen to be equal to the maximum  of the surface
gravities of $(\Sigma,g,V)$. On the other hand, if $\pSo = \emptyset$, 
the mass of the RS will be defined suitably below, in the proof of theorem
\ref{T1.1}. We only consider RS with mass $m_{0}$ in the range \eq{rngm}
(if $\pSo \ne \emptyset$, this property follows from the restriction
\eq{kapineq} on $\kappa$). It should be stressed that we are {\it not}
comparing  manifolds
and/or metrics, but we are only using the resulting scalar functions $V$ and
$W$.

Let $r(\cdot)$ be the function $V_0 \to r(V_0)$ constructed at the end of 
Section~\ref{SgK}. Composing $r$ with $V$ we obtain functions $r(V(\cdot))$ 
and  $W_{0}(r(V(\cdot))$ defined on $\Sigma$. By an abuse of notation we
shall still denote those functions by $r$ and $W_{0}$. 

In the same manner, we can define a RS from other 
solutions with the property that $W$ is a function of $V$ only. (In Lemma 
\ref{Lrigidity} below we will also include the Nariai case).

Following \cite{BeigSimon3} we define $\psi(V)$ to be that unique solution of
the equation
\begin{equation}
\label{dPsi}
\psi^{-1} \frac{d\psi}{dV}  =  - V W_0^{-1} \frac{m_0}{r^3}
\end{equation}
which goes\footnote{Using the asymptotic behavior of $V(r)$ and $r(V)$
  it is not too difficult to show that solutions of \eq{dPsi} are
  uniformly bounded on $[0,\infty)$, and approach a non-zero constant
  at infinity unless identically vanishing. Since solutions of
  \eq{dPsi} are defined up to a multiplicative constant, we can choose
  this constant so that our normalization holds.} to $1$ as $V$ goes
to $\infty$. (In particular $\psi\equiv 1$ when $m_0=0$.)
Here $r=r(V)$ is again the function defined at
the end of Section~\ref{SgK}. Standard results on ODE's show that
solutions of \eq{dPsi} have no zeros unless identically vanishing, and
that
$$\Psi\equiv\psi\circ V$$ can be extended by continuity to a smooth
function on
$\overline\Sigma$, still denoted by $\Psi$, which satisfies
  $$\Psi>0\;, \quad \Psi|_{\pSinfty}=1 \; .$$ We also define
\begin{eqnarray}
\widetilde g_{ij} &= &V^{-2}\Psi^{4}g_{ij}\;,
\nonumber \\ 
\widetilde W &=& \Psi^{-4} W\;,
\nonumber \\ 
\widetilde W_{0} &=& \Psi^{-4} W_{0}\;.
\label{gw}
\end{eqnarray}

We proceed with a computation which is required in Lemma \ref{LHm} 
as well as in Lemma \ref{Lahor}.
Consider a level set $\{V= \mbox{const}\}$ of $V$ which is a
smooth hypersurface in $\overline{\Sigma}$, with unit normal $n_{i}$,
induced metric $h_{ij}$, scalar curvature ${}^{2}{\cal R}$, second
fundamental form $p_{ij}$ defined with respect to an inner pointing
normal, mean curvature $p = h^{ij}p_{ij}$; we denote by $q_{ij}$ the
trace-free part of $p_{ij}$: $q_{ij} = p_{ij} - \frac{1}{2}h_{ij}p$.
Using Equation~\eq{W0}, the Equation (\ref{f1}) with $g=g_0$ and
$V=V_0$, together with the relation
\begin{equation}
  \label{Vder}
  \frac{d V_0}{d r} = \frac{\sqrt{W_0}}{V_0} 
\end{equation}
 we obtain
\begin{equation}
\label{dW0}
V^{-1} \frac{d W_{0}}{dV} = -\frac{2}3 \Lambda- \frac{4m_0}{r^{3}}\; .
\end{equation} 
To obtain~\eq{dWtil} we use, in this order, the definitions~\eq{gw}, 
the Equations \eq{f1}--\eq{f2}, Equations~\eq{dW0} and \eq{dPsi},
 and the Codazzi-Mainardi equation: 
\begin{eqnarray}
\label{dWtil}
\lefteqn{V^{-1} \widetilde W^{-1} D^{i}V D_{i}(\widetilde W -
 \widetilde W_0) } 
 \nonumber \\
 &  & {} =  V^{-1} W^{-1}D^{i}V (D_{i}W)  - V^{-1}\frac{dW_0}{dV} -
4 V^{-1} \Psi^{-1} \frac{d\Psi}{dV} (W - W_0)  \nonumber \\
 & & {} =  (2 { R}_{ij} - { R}g_{ij}) n^{i}n^{j} +
\frac{2}{3} \Lambda + \frac{4m_0}{r^3}  - \frac{4m_0}{r^3}(1 - W_0^{-1} W)
\nonumber\\
& & {} = - {}^{2}{\cal R} - q_{ij}q^{ij} + \frac{1}{2}p^{2} +
\frac{2}{3} \Lambda + \frac{4m_0}{r^3} - \frac{4m_0}{r^3}(1 - W_0^{-1}
W) \;.
\end{eqnarray} 

\begin{Lemma}
  \label{LHm}
   Under the conditions of Theorem~\ref{Ttop}, suppose further that
   the scalar curvature $R'$ of the metric $g'=V^{-2}g$ is constant on
   $\pSinfty$. Let $V$ be normalized so that 
  \eq{Rprime0} holds, with $A'_\infty=4\pi\ell^2$ when $\pSinfty=T^2$. If $m$
  is the \GHn\ mass parameter defined as in \eq{masspar}, then
  \begin{equation}
\label{Hm}
  \int_{\pSinfty}D'_{i}(\widetilde W - \widetilde W_0) dS'^{i} = -
  \left(\frac {2\Lambda} 3\right)^2A'_{\pSinfty}(m - m_0)\;,
\end{equation}
where $dS'^{i}$ denotes the outer-oriented area element of the metric
$g'=V^{-2}g$, and $A'_{\pSinfty}$ is the area of ${\pSinfty}$ with
respect to that metric.
\end{Lemma}

\proof 
Using 
\begin{eqnarray}
  \label{Wtrans}
  D'^i (\widetilde W - \widetilde W_0)n'_i & = & \frac
  1{\sqrt{W'}}D_{i}(\widetilde W - 
  \widetilde W_0) D^i V\;
\end{eqnarray}
and ~\eq{dWtil}, the left hand side of \eq{Hm} reads
\begin{equation}
\label{intWtil}
\int_{\pSinfty}
  \frac{V\widetilde W}{\sqrt{W'}} 
\left[ - {}^{2}{\cal R} - q_{ij}q^{ij} + \frac{1}{2}p^{2} + \frac{2}{3} \Lambda + 
\frac{4m_0}{r^3} - \frac{4m_0}{r^3}(1 - W_0^{-1} W)\right] d^2\mu_{g'}
\end{equation} 
where $d^2\mu_{g'}$ is the two-dimensional surface measure associated with
the metric $g'$. Chasing through the definitions one finds that
\begin{equation}
  \label{factor}
  \frac{V\widetilde W}{\sqrt{W'}} \approx 
\sqrt{-\frac \Lambda      3} V ^3
\end{equation}
near $\pSinfty$. {}From the definition of $V_0$ we further have
$$r \approx \sqrt{-\frac{3}\Lambda }V\;,$$ again near $\pSinfty$, so that
$\lim_{V\to\infty}V \widetilde W/(\sqrt{W'}r^3)= (-\Lambda/3)^2$. It
follows that the second to last term in \eq{intWtil} gives a
contribution
\begin{equation}
  \label{mcont}
  \left(\frac {2\Lambda} 3\right)^2A'_{\pSinfty}m_0
\end{equation}
where $A'_{\pSinfty}$ denotes the $g'$-area of  the connected component 
of $\pSinfty$ under consideration.  Equation \eq{Rprime} and its equivalent 
with $W$ replaced by $W_0$ show that $$(1 - W_0^{-1} W)\to_{V\to\infty} 0\;$$ 
so that the last term drops out from \eq{intWtil}. Furthermore, by 
Equation~\eq{qfall2} we have
$$\frac{V \widetilde W}{\sqrt{W'}}q_{ij}q^{ij}
=O(V^{-3})\to_{V\to\infty}0\;,$$ and it remains to analyze the
contribution of $-V \widetilde W \left( {}^{2}{\cal R}-
  \frac{1}{2}p^{2} - \frac{2}{3} \Lambda\right)/\sqrt{W'}$ to the
integral \eq{Hm}. To do this, note that
\begin{eqnarray*}
  A_{1/\epsilon}\equiv A(\{V= 1/\epsilon\}) & = & \int_{V'=\epsilon}
  d^2\mu_{g} \\ & = & \int_{V'=\epsilon} V^2 d^2\mu_{g'}\;\approx\;
  \epsilon^{-2} A'_{\pSinfty}\;,
\end{eqnarray*}
where $d^2\mu_g$ 
is the induced measure on $\pSinfty$ associated with the metric $g$.
It follows that
\begin{eqnarray}
  \lefteqn{ - \int_{V'=\epsilon}\frac{V \widetilde W}{\sqrt{W'}}\left(
      {}^{2}{\cal R}- \frac{1}{2}p^{2} - \frac{2}{3} \Lambda\right)
    d^2\mu_{g'}\phantom{xxxxxxxxxxxxxxxxxxxxxxxxxxx}} \nonumber & & \\ 
  & \approx \phantom{{\epsilon 0} }& - \sqrt{-\frac \Lambda 3}\frac{
    1}{ \epsilon} \int_{V'=\epsilon}\left( {}^{2}{\cal R}-
    \frac{1}{2}p^{2} - \frac{2}{3} \Lambda\right) d^2\mu_{g}
  \nonumber\\ & \approx \phantom{{\epsilon 0} }& - \sqrt{-\frac \Lambda
    3}\sqrt{\frac{A_{1/\epsilon}}{A'_{\pSinfty}}} \int_{V'=\epsilon}\left(
    {}^{2}{\cal R}- \frac{1}{2}p^{2} - \frac{2}{3} \Lambda\right)
  d^2\mu_{g} \nonumber\\ & \to_{\epsilon\to 0} & \label{m} -\left(\frac
    {2\Lambda} 3\right)^2A'_{\pSinfty}m\;,
\end{eqnarray}
where 
\begin{equation}
  \label{m?}
  m\equiv{\lim}_{
    \epsilon \to 0} \frac 14 \left(-\frac{\Lambda
      A'_{\pSinfty}}{3}\right)^{-3/2}\sqrt{A_{1/\epsilon}}
  \int_{\{V=1/\epsilon\}}( {}^{2}{\cal R} - \frac{1}{2}p^{2}
  -\frac{2}{3}\Lambda) dA \; .
\end{equation}
To finish the proof we need to show that $m$ in \eq{m?} is indeed the
\GHn\ mass as defined in Equation~\eq{masspar}.
In the torus case this follows immediately from the normalization
condition  $A'_\infty = 4\pi\ell^2$; for the remaining topologies this can be
seen as follows: if $V$ is normalized so
that \eq{Rprime0} holds, then \eq{RR} implies
$$ {}^{2}{\cal R'}\Big|_{x=0}= - \frac 23 \Lambda k\;.$$ When
$g_\infty\ne 1$ the Gauss--Bonnet theorem gives
\begin{eqnarray*}
  8\pi |1-g_\infty| & = & \left|\int {}^{2}{\cal R'}
  d^2\mu_{g'}\right|
\\ & = & - \frac 23 \Lambda A'_{\pSinfty}\;,
\end{eqnarray*}
which shows that the mass defined by Equation~\eq{m?} coincides with
that of \eq{masspar}.
 \qed

As to the subsequent Lemma, recall from Theorem \ref{T1.2} 
that $\partial_{1}\Sigma$ refers to the the component of $\partial\Sigma$ 
of the given solution with the largest surface gravity.

\begin{Lemma}
\label{Lahor}
   Under the conditions of Theorem~\ref{Ttop}, we have  
  \begin{equation}
\label{Ahor1}
  \int_{\partial_{1}\Sigma} 
  {\widetilde W}^{-1/2}
   \widetilde D_{i}(\widetilde W - \widetilde W_0) d\widetilde S^{i} = 
  8\pi \left[ \left(g_{\partial_{1}\Sigma} - 1 \right) - 
  \frac{A_{\partial_{1}\Sigma}}{A_0}\left(g_{\infty} - 1 \right) \right]
\end{equation}

\end{Lemma}
\proof
We integrate \eq{dWtil} over $\partial_{1}\Sigma$. We note that Equation~\eq{f2}
multiplied by $V$ and contracted with two vectors tangent to $\pSo $
shows that $\pSo $ is totally geodesic; equivalently, $q_{ij} = 0$.
We introduce $${}^{2}{\cal R}_0 = \frac{2}{3} \Lambda +
\frac{4m_0}{r_0^3} \;,$$ 
the scalar curvature of the metric $d\Omegak^2$.
Using  \eq{dWtil} and the Gauss--Bonnet theorem, the left hand side
of \eq{Ahor1}  can be written as
\begin{equation}
  \label{Ahor2}
  \int_{\partial_{1}\Sigma} (-{}^{2}{\cal R} +\frac{2}{3} \Lambda +
\frac{4m_0}{r_0^3}) dA =  
 \int_{\partial_{1}\Sigma} (-{}^{2}{\cal R} + {}^{2}{\cal R}_0)dA =
  8\pi \left(g_{\partial_{1}\Sigma} - 1 \right) + {}^{2}{\cal R}_0 A_{\pSm}
 \label{Isr} 
\end{equation}
Equation \eq{Ahor1} is then obtained by  eliminating ${}^{2}{\cal R}_0$  from \eq{Ahor2}, 
using the Gauss--Bonnet theorem 
for the \gK\ metrics:
$$8 \pi (1-g_{\infty}) = {}^{2}{\cal R}_0 A_{0}\;.$$
\qed

The following elliptic equation for $\widetilde W - \widetilde W_0$ will be the 
crucial ingredient in the proof of the theorems. It is also  useful for
Lemma \ref{Lrigidity}.

\begin{eqnarray}
\label{lw}
\lefteqn{(\widetilde \Delta - a)(\widetilde W - \widetilde W_{0}) = }
\nonumber\\
& &  = \frac{1}{4} \widetilde W^{-1} \widetilde R_{ijk} \widetilde R^{ijk}
+ \frac{3}{4} \widetilde W^{-1} 
\widetilde D_{i}(\widetilde W - \widetilde W_{0})
\widetilde D^{i}(\widetilde W - \widetilde W_{0})\ , 
\end{eqnarray}
with \begin{equation}
 \label{a}
a =   \frac{5}{3 r^3}m_0\Lambda V^{4}  W_{0}^{-2}~ \widetilde W\;, 
\end{equation}
$\widetilde \Delta$ being the Laplace operator of the metric
$\widetilde g_{ij}$, and $\widetilde R_{ijk}$  --- the Cotton tensor 
of $\widetilde g_{ij}$. 
This equation is obtained by specializing\footnote{The assumption of
  spherical symmetry of the level sets of the reference solution made
  in   \cite{BeigSimon3}  is not needed to obtain Equation \eq{a}.} Equation~(5.4) of
\cite{BeigSimon3} (which  has been used in that paper in the context
of a uniqueness proof for static  perfect fluid solutions) to the
present case with $8\pi\rho = -8\pi p = \Lambda$. 

It is important to stress that Equation~\eq{lw}, as it stands, makes
only sense on the set $\{dV\ne 0\}$, because of the factors
$\widetilde W^{-1}$ appearing there. However, it follows from Equation
\eq{f1} that the set $\{dV=0\}$ has no interior: indeed, if $dV$
vanishes on a connected open set then $V$ is constant there, which is
compatible with Equation~\eq{f2} only if $V$ vanishes there. This
contradicts our hypothesis that $V$ vanishes only\footnote{Let us
  mention that if $V$ is zero on an open 
  set, then the Aronszajn unique continuation theorem~\cite{aronszajn}
  shows in any case that $V$ must be identically zero on $\Sigma$.} on
$\pSo $ .  Hence Equation~\eq{lw} holds on an open dense set of
$\Sigma$. Since the left hand side of Equation~\eq{lw} is a smooth
function on $\Sigma\setminus\pSo $, the right hand side thereof is
smoothly extendible by continuity to a smooth function on
$\Sigma\setminus\pSo $, and Equation~\eq{lw} holds everywhere on this
set with the right hand side being understood in the sense explained
here.

\begin{Lemma}
  \label{Lrigidity} Let $\Lambda \in \R$, and let $(\Sigma,g,V)$ be a
  solution of  \eq{f0}--\eq{f2} such that
\renewcommand{\theenumi}{\alph{enumi}}
  \begin{enumerate}
  \item \label{one} either $W\equiv W_0$ for $W_0$ defined from the
generalized Kottler or from the Nariai solution \eq{Nar}, or
  \item \label{two} $(\Sigma,g)$ is locally conformally flat.
  \end{enumerate}
  Suppose further that $\Sigma$ is a union of compact
  boundary-less level sets of $V$.
  Then:
\renewcommand{\theenumi}{\arabic{enumi}}
\renewcommand{\theenumii}{\roman{enumii}}
  \begin{enumerate} \item \label{oner} Every connected component $\cV$
    of the set $\{p\in\Sigma\;|\; dV(p)\ne 0\}$ ``corresponds to''
    one of the \gK\ solutions \eq{Kot}, or to  one
    of the generalized Nariai solutions
    \eq{Nar}, or is flat.  More precisely, there exists
    an interval $J\subset\R$, a two-dimensional compact Riemannian manifold
    $({}^2M,d\Omegak^2)$,   with $d\Omegak^2$  an (r-independent) metric   of
   constant Gauss curvature $k = 0, \pm 1$, and a diffeomorphism
    $\psi:\cV\to J\times{}^2M$ such that, transporting $g$ and $V$ to
    $J\times{}^2M$ using $\psi$, we have:
    \begin{enumerate}
    \item \label{PKot}Either there exists a constant $\lambda>0$ such that
      $V=\lambda V_0$ and
\begin{eqnarray}
    \label{Kotstd}
  & g  =  V_0^{-2}dr^2 + r^2 d\Omegak^2\;,\ r\in J\;,&
\\  &V_0^2  =  k -
    \frac {2m}r - \frac \Lambda 3 r^2\;, &
   \end{eqnarray} 
 \item \label{PNar}or, when $k\Lambda>0$, there exists a constant
   $\lambda\in\R$ ($\lambda >0$ if $\Lambda >0$) such that
\begin{eqnarray}
    \label{Narstd}
    &g  =  V^{-2}dz^2 + |\Lambda|^{-1} d\Omegak^2\;,\ z\in J\;, &
\\ 
   \quad & V^2  =  \lambda -  \Lambda  z^2
\;,
  \end{eqnarray} 
\item \label{Pflat}or, when $k=\Lambda=0$, there exists a constant $\lambda>0$ such
  that $V=\lambda z$ and 
\begin{eqnarray}
    \label{flat}
   &g  =   dz^2 +  d\Omegak^2\;,\ z\in J\;.&
   \end{eqnarray} 
    \end{enumerate} 
 (In each case the interval $J$ is constrained by the condition that
 $V$ and $V^2$ be  non-negative).

\item \label{twor} Under condition~\ref{one}.~above, if $\Sigma$ is
  connected and if $W_0$ (considered as a function of $V$) has no
  zeros in the interval where $V$ takes its values, 
  \begin{equation}
    \label{Vnocrit}
    \forall \;p\in\Sigma \qquad W_0(V(p))\ne 0 \;,
  \end{equation}
  then $\cV=\Sigma$, thus Equations \eq{Narstd} or
  \eq{Kotstd}  hold globally on $\Sigma$.
  \end{enumerate}
\end{Lemma}

\noindent{\sc Remarks:} 1. Here we do not make any hypotheses on the sign of
$\Lambda$. 

2. The result  is local, in particular it is
sufficient to be able to invert $r_0(V_0)$ locally on the range of the
values of $V$ under consideration to obtain $W_0(V)$. 

3. The set $(\Sigma,g,V)$ corresponding to the 
metric \eq{flat} arises from a boost Killing vector in suitably
identified Minkowski space-time.

4. We note that the set $\cV$ could be empty, as is the case for
$\R\times T^3$ with the obvious flat metric. Our analysis does not say 
anything about the metric on regions where $dV$ vanishes.

5. We note that the \gK\ and the generalized Nariai
metrics also 
arise naturally in the generalized Birkhoff theorem, see
\cite{Goenner2,ForMac}, and also \cite{Stanciulescu} for a very clear
treatment in the $\Lambda>0$ case.

6. The Lemma can easily be reformulated by taking any conformally flat
solution of \eq{f1} and \eq{f2} as a reference solution. The condition 
of conformal flatness is required to ensure that \eq{lw} holds and 
excludes, in particular, the Horowitz-Myers solutions with toroidal $\scrp$ 
\cite{HorowitzMyers} as RS.
\medskip

\proof The proof is an adaptation of an argument of
\cite{Chstaticelvac} to the current setting. Suppose that $W=W_0$ for
some $W_0$; Equation~\eq{lw} shows then that $\widetilde {
  R}_{ijk} \widetilde { R}^{ijk}$ vanishes, so that $(\Sigma,g)$
is locally conformally flat. It then follows that condition\emph{~\ref{two}}.~holds in both cases.

We start by removing from $\Sigma$ some undesirable points: set
\begin{eqnarray*}
  \Sigma_{\mbox{\scriptsize{sing}}} & \equiv & \{p\in \Sigma|\; \mbox{the
  connected component of the set $\{q|V(q)=V(p)\}$ }
\\ & & \quad \mbox{containing $p$
  contains a point $r$ such that $dV(r)=0$. }\}\;,
\\ \Sigma'& \equiv & \Sigma \setminus \Sigma_{\mbox{\scriptsize{sing}}}\;.
\end{eqnarray*}
$\Sigma_{\mbox{\scriptsize{sing}}}$ is a closed subset of $\Sigma$, so
that $\Sigma'$ is still a manifold. It follows from Sard's theorem
that $\Sigma'\ne\emptyset$. We note that $\Sigma'$ still satisfies all
the hypotheses of the Lemma, except perhaps for being
connected. By construction all the level sets of 
$V$ are non-critical in $\Sigma'$. (Recall that a level set $\{V=c\}$
of $V$ is non-critical if $dV$ is nowhere vanishing on $\{V=c\}$.)

Let $\cU$ to be any connected component of $\Sigma'$.
Compactness of the level sets of $V$ implies\footnote{The possibility 
  that $\cU$ is diffeomorphic to $S^1\times {}^2M$ (or some twisted
  version thereof) is excluded by the
  fact that $dV$ does not vanish on $\cU$.} that $\cU$ is
diffeomorphic to $I\times {}^2M$, for some two-dimensional compact
connected manifold ${}^2M$ and some interval $I\subset \R$, with $V$
equal to $c$ on $\{c\}\times{}^2M$, $c\in I$, and that on $\cU $ the
function $V$ can be used as a coordinate. Further we can introduce on
${}^2M$ a finite number of coordinate patches with coordinates $x^A$,
$A=1,2$, so that on $\cU$ the metric takes the form
\begin{equation}
  \label{gcor}
  g=W^{-1} dV^2 + h_{AB}dx^A dx^B \ .
\end{equation}
Let, as before, $q_{AB}dx^A dx^B$ be the trace free part of the
extrinsic curvature tensor of the level sets of $V$ --- in the
coordinate system of~\eq{gcor}
\begin{equation}
  \label{lam}
  q_{AB} = \sqrt{W}\Big( \frac{\partial h_{AB}}{\partial V}
  -\frac{1}{2}h^{CD}\frac{\partial h_{CD}}{\partial
    V}h_{AB}\Big) \ .
\end{equation}
Equations~\eq{lam} and~\eq{LL} imply that $q_{AB}$ vanishes hence
$\frac{\partial h_{AB}}{\partial V}$ is pure trace, that $W=W(V)$, and
that $\det \gamma_{AB}$ is a product of a function of $V$ with a
function of the remaining coordinates. We thus have
\begin{equation}
  \label{gcor2}
  h=W(V)^{-1} dV^2 + r(V)^2 d\Omega^2\ .
\end{equation}
for some positive function $r(V)$, where $d\Omega^2$ is a $V$-independent
metric on ${}^2M$.  Next, from~\eq{f2} and from the Codazzi-Mainardi
equations~\eq{CM1a}, respectively~\eq{CMnn}, applied to ${}^{2}M
\subset {\cU}$, we find that the mean curvature $p$ of all level surfaces,
respectively their Ricci scalars, are constant. Hence
$({}^{2}M,d\Omega^2)$ is a  
space of constant curvature, and scaling $r$ appropriately we can
without loss of generality assume that the Gauss curvature $k$ of the
metric $d\Omega^2$ equals $0,\pm1$, as appropriate to the genus of ${}^2M$.
We define 
\begin{equation}
\label{Ldef}
L = \frac{dW}{dV} + 2\Lambda V\;.
\end{equation}
Evaluating  \eq{f1} for the metric \eq{gcor2}, we find 
\begin{equation}
\label{drdV}
\frac{dr}{dV} = - \frac{rL}{4W}\;.
\end{equation}
Equations~\eq{f1}--\eq{f2} for the metric 
\eq{gcor2} are equivalent to \eq{Ldef}--\eq{drdV} together with
\begin{eqnarray}\label{dLdV0}
&2W\left(\Lambda -\frac{k}{r^2}\right) = L\left(V^{-1}W  - \frac L 8
\right)\;,& \\
\label{dLdV}
&W\frac{dL}{dV} = \frac{3}{4} L^2 + (V^{-1}W - \Lambda V) L \;.&
\end{eqnarray}
 These equations arise \emph{e.g.} by adapting Equations (3.16) and (3.17) 
of \cite{BeigSimon3} to the present case (namely by setting $ 8\pi\rho = - 8\pi p = \Lambda$,
$L_{1} = L$ and  $C^{2} = k$, and allowing the constant $k$ to take
negative values). 
Suppose, first, that there exists $V_*$ such that $L(V_*)=0$. Equation
\eq{dLdV}  shows then that  $L \equiv 0$, and from \eq{dLdV0} one
obtains
\begin{equation}
  \label{eq:klam}
\Lambda = \frac k {r^2}\;.
\end{equation}
If $k=0$ then $\Lambda$  vanishes as well; further $r$ is 
constant by Equation~\eq{drdV} and can therefore be absorbed into $d\Omega^2$.
Integrating Equation~\eq{Ldef} one finds that there exists a strictly positive
constant $\lambda$ such that $W=\lambda^2$, defining a coordinate
$z$ by the equation $z=V/\lambda$ proves  point\emph{~\ref{Pflat}}.~on 
$\cU$.
Next, if $k\ne 0$ Equation~\eq{eq:klam} gives $k\Lambda >0$ as
desired, together with $r^2=-1/|\Lambda|$. Integrating
Equation~\eq{Ldef} one obtains 
$$W=\Lambda(\lambda-V^2)\;,$$ for some constant $\lambda\in
\R$. Introducing the  
coordinate $z$ via the equation $V^2=\lambda-\Lambda z^2$ establishes
point\emph{~\ref{PNar}}.~on $\cU$. 

In the case of $L$ without zeros we obtain, from \eq{Ldef}, \eq{drdV} 
and \eq{dLdV}, that $$\frac{d}{dV}\left( \frac{V\sqrt{W}}{rL}\right) = 0\;,$$
which implies that there exists a non-vanishing constant $\alpha$ such that
\begin{equation}
\label{LW}
L  =  \alpha V\frac{\sqrt{W}}{r}\;.
\end{equation}
Using \eq{drdV} one is led to
\begin{equation}
\label{dVW}
\frac{dV}{dr}  =  -\frac{4\sqrt{W}}{\alpha V}\;.
\end{equation}
Next we define 
\begin{equation}
\label{mdef}
m(V) = -\frac \alpha 4 r^2 \sqrt{W} + \frac{\Lambda r^3}{3}\;;
\end{equation}
{}from \eq{Ldef}, \eq{LW} and \eq{dVW} we obtain $dm/dV = 0$,
\emph{i.e.}  $m$ is a constant. Equation~\eq{dLdV0} gives
$$ V^2 = \frac {16}{\alpha^2}\left( k -
    \frac {2m}r - \frac \Lambda 3 r^2\right)\;.$$ Equation~\eq{drdV}
  shows that we 
can use $r$ as a coordinate, and Equation~\eq{dVW} implies that the metric is of the 
desired form \eq{Kotstd}. This establishes point\emph{~\ref{PKot}}.~on $\cU$. 

Let $\cV$ be the connected component of $\{dV\ne 0\}\subset \Sigma$
that contains $\cU$. To establish point\emph{~\ref{oner}}.~of the Lemma we
need to show that $\cV=\cU$. 
We claim that $\cU$ is open in $\cV$ --- and hence in $\Sigma$ ---
which can be seen as follows: Let $p\in \cU$, we thus have $dV(q)\ne
0$ for all $q$ such that $V(p)=V(q)$. By Equation~\eq{gcor2}
$|dV|_{g}=\sqrt{W}$ is constant on the intersection with $\cU$ of the
level set $V^{-1}(V(p))$ of $V$ through $p$, so that
$$
\inf_{V^{-1}(V(p))\cap\cU}|dV|_{g} 
> 0 \ ,
$$ which easily implies that all nearby level sets in
$\cU\subset\Sigma'$ are non-critical.

Let us show now that $\cU$ is closed in $\cV$. To see that, consider a
sequence $p_i\in\cU$ such that $p_i\to p\in \cV$. By definition of
$\cV$ the function $|dV|_g$ has no zeros on $\cV$, hence $dV(p)\ne 0$.
Now it follows from \eq{LL} that $|dV|_g$ is locally constant on
smooth subsets of level sets of $V$, which easily implies a) that the
connected component of $V^{-1}(V(p))$ containing $p$ is smooth and b)
that $|dV|_g$ is nowhere vanishing there. Compactness of the level
sets of $V$ implies that all the connected components of level sets
intersecting a neighborhood of $p$ are non-critical, and hence are in
$\Sigma'$. It then follows that $p\in\cU$.

We have thus shown that $\cU$ is both open and closed in $\cV$;
connectedness of $\cV$ shows that $\cU=\cV$, and point\emph{~\ref{oner}}.~is
established. 

To prove part\emph{~\ref{twor}.}, we note that the equality $W(p)=W_0(V(p))$
together with Equation~\eq{Vnocrit} shows that $V$ has no critical
points on $\Sigma$; as $\Sigma$ is connected the set $\cV$ of
point~\emph{\ref{oner}}.~coincides with $\Sigma$, and
point\emph{~\ref{twor}}.~follows from point\emph{~\ref{oner}}.  \qed

\subsection{Proofs}

{\sc Proof of Theorem \protect\ref{T1.1}:}
Suppose that $\pSo =\emptyset$. We first consider as RS a generalized
Kottler solution with $m_0 = 0$
 (see Equation~\eq{krel}). This leads to 
\begin{equation}
  \label{psiandV}
  \Psi \equiv 1\;, \qquad
\widetilde W_{0}(V_0)= - \frac{\Lambda }{3}(V^2_{0}-k)\;.
\end{equation}
We further normalize $V$ as in Proposition~\ref{P2}, so that by
\eq{Rprime}, \eq{Rprime0.0} and \eq{Rprime0} we have
$$\widetilde W - \widetilde W_{0}\to_{r\to\infty}0\;. $$  (Actually
when $\pSinfty=T^2$, the 
normalization of $V$ does not play any role, as we make claims only
about the sign of $m$ in this case.) Equation
\eq{lw} together with the maximum principle shows that
\begin{eqnarray}
 & \label{rp1}
  \widetilde W -
\widetilde W_{0}\le 0 \;\;\mbox{on}\;\; \Sigma\;, 
& \\ & \label{rp2}
 n^{\prime\, i}D_{i}'(\widetilde W - \widetilde W_{0})|_{\pSinfty}\ge 0\;,
\end{eqnarray}
where $n'$ is the \emph{outer} pointing $g'$-unit normal to
$\pSinfty$.   Further, equality is attained in
\eq{rp1} or in \eq{rp2} if and only if $W\equiv
W_0$~\cite[Theorems~3.5 and 3.6]{GT}. 
Thus Lemma~\ref{LHm} together with Equation~\eq{rp2} shows that
$$m\le 0\;.$$ Assume now that $m=0$ in the case $\pSinfty = S^2 $; as
an indirect argument, we also assume that $m=0$ in the  $T^2$ case, or
that $m\ge m_{crit}$ in the remaining cases. In the sphere or torus
case from the strong maximum principle we obtain
\begin{equation}
  \label{Wident}
  W \equiv W_{0}\;.
\end{equation}
In the higher genus cases we consider (\ref{lw}) again but take here as RS
a \gK\ solution with the same mass as the given one, $m_0 = m$. 
Equations~\eq{rp1}--\eq{rp2} hold again; then
Lemma~\ref{LHm} shows that equality must hold in \eq{rp2}.
Applying the maximum principle again yields Equation \eq{Wident}.
We note that both point~\emph{\ref{one}}.~as well as the structural hypotheses of
Lemma~\ref{Lrigidity} hold under the hypotheses of Theorem~\ref{T1.1}.
Equation~\eq{Wident} and the discussion of Section~\ref{SgK} show that
point\emph{~\ref{twor}}.~of that Lemma applies, so that the given solution
must be a member of the \gK\ family with $m$ in the range \eq{rngm}
(the generalized Nariai metrics are excluded as they do not satisfy
the asymptotic hypotheses of Theorem~\ref{T1.1}).
In the case $\pSinfty=S^2$ point~\emph{\ref{p1t1}}.~readily follows.
 In the remaining cases none of these solutions
has the topology required in Theorem~\ref{T1.1}, which gives a
contradiction and establishes Theorem~\ref{T1.1}.
\qed

\noindent{\sc Proof of Theorem \protect\ref{T1.2}:}  
By choice of the RS we have 
$(\widetilde W - \widetilde W_0)\vert_{\pSo } = 0$. We normalize $V$
again so that $\mbox{lim}_{\rightarrow \infty} (\widetilde W -
\widetilde W_0) = 0$ holds, \emph{cf.\/} Proposition~\ref{P2} and
equation~\eq{Rprime}.  Negativity of $m_0$ implies that $a$ in
(\ref{lw}) is nonnegative.  The maximum principle applied to Equation
(\ref{lw}) gives $\widetilde W - \widetilde W_0 \le 0$ on $\Sigma$,
with equality being achieved somewhere if and only if $W\equiv W_0$.
Moreover, as in the proof of part 2, the boundary version of the
strong maximum principle~\cite[Theorem~3.6]{GT} implies that 
$n^{i\,\prime} D'_{i}(\widetilde W - \widetilde W_0) > 0$ on $\pSinfty$ 
unless $W = W_{0}$.  Lemma~\ref{LHm} allows us to conclude that either 
$m < m_0$ or $W \equiv W_{0}$. In that last case point~\emph{\ref{twor}}.~of
Lemma~\ref{Lrigidity} implies that $(\Sigma,g,V)$ corresponds to a
\gK\  solution. In any case there holds $m \le m_0$.

To prove the area inequality in \eq{ourineq} requires some care as 
the metric $\widetilde g$ defined in Equation \eq{gw} is singular at 
$\Sigma$, so that standard maximum principle arguments such 
as~\cite[Theorem~3.6]{GT} do not apply. We proceed as follows.
 By choice of $W_0$ we have $\widetilde W=\widetilde W_0$ on $\pSm $. 
Further, Equation~\eq{Wvanish} shows that $n^iD_i(\widetilde
W-\widetilde W_0)$ vanishes there.  
De l'Hospital's rule, the non-vanishing of $dV$ at $\pSo $, 
and the requirement $\widetilde W-\widetilde W_0\le 0$ lead to 
$$n^j n^i D_i D_j(\widetilde W-\widetilde W_0)\Big|_{\pSo } = 
\lim_{V\to0} \frac{D^i  V D_i (\widetilde W-\widetilde W_0)}{V} \le 0\;.$$
It follows that the left-hand-side of Equation~\eq{Ahor1} is
non-positive, which establishes the second part
of \eq{ourineq}. \qed

\noindent{\sc Proof of Corollary \protect\ref{C1.2}:} Assume that
$\pSo$ is connected and 
that \eq{rds1x} holds; we want to show that \eq{ourineq} implies an
inequality inverse to \eq{rds1x}. In order to do this, note first that
by \eq{ourineq} the mass $m$ is non-positive, and Equation \eq{rds1x}
implies that $g_{\pSo } > 1$. It is useful to introduce a
genus-rescaled area radius $\rds$ by the formula
$$\rds = \sqrt{\frac{A_{\pSo }}{4\pi(g_{\pSo } - 1)}}\;.$$  In terms
of this object, the inequality \eq{rds1x} reads 
\begin{equation}\label{iPi}
  2m |g_\infty -1|^{3/2}+ \left(\rds +
    \frac{\Lambda}{3}\rds^3\right)|g_{\pSo} -1|^{3/2} \ge 0 \;,
 \end{equation} 
 It follows that $\rds + \frac{\Lambda}{3}\rds^3 \ge 0$, and the
 Galloway--Schleich--Witt--Woolgar inequality $g_{\pSo}\le g_{\infty}$
 implies
\begin{equation}\label{iPix}
  2m + \rds + \frac{\Lambda}{3}\rds^3 \ge 0 \;,
 \end{equation} 
 Let us denote by $r_0$ the $\rds$ corresponding to the relevant \gK\ 
 solution:
 $$r_0 = \sqrt{\frac{A_0}{4\pi(g_{\pSinfty } - 1)}}\;.$$ The
 inequality \eq{iPix} is actually an equality for the \gK\ solutions,
 therefore it holds that
 $$ 2m_0 + r_0 + \frac{\Lambda}{3}r_0^3 = 0 \;.$$ We have $r_0 \ge
 1/\sqrt{-\Lambda}$ from \eq{rngr}, and $m \le m_0$ , $\rds \ge r_{0}$
 from \eq{ourineq}, so that
\begin{eqnarray}
\lefteqn{2m + \rds + \frac{\Lambda}{3}\rds^3 = 
2m + \rds + \frac{\Lambda}{3}\rds^3 -2m_0 - r_0 - \frac{\Lambda}{3}r_0^3 =
{}} \nonumber \\ 
& & {} = 2(m - m_0) + (\rds - r_0)[1 + \frac{\Lambda}{3}(\rds^2 + \rds r_0 +
r_0^2)] \le {} 
\nonumber\\
& & {} \le (\rds - r_0)(1 + \Lambda r_0^2) \le 0\;. \label{finalgpi}
\end{eqnarray}
It follows from Equations~\eq{iPix}--\eq{finalgpi} that $\rds = r_0$,
$m=m_0$, and the rigidity part of Theorem~\ref{T1.2} establishes
Corollary~\ref{C1.2}.\qed

\def\cprime{$'$}
\providecommand{\bysame}{\leavevmode\hbox to3em{\hrulefill}\thinspace}


\begin{thebibliography}{10}

\bibitem{AbbottDeser}
L.F. Abbott and S.~Deser, \emph{Stability of gravity with a cosmological
  constant}, Nucl.\ Phys. \textbf{B195} (1982), 76--96.

\bibitem{manderson:stationary}
M.T. Anderson, \emph{On stationary vacuum solutions to the {E}instein
  equations}, {URL} {\url{http://www.math.sunysb.edu/~anderson},}
  gr-qc/0001091, to appear in {Annales Henri Poincar\'e}, 2000.

\bibitem{manderson:static}
\bysame, \emph{On the structure of solutions to the static vacuum {E}instein
  equations}, {{URL}} \url{http://www.math.sunysb.edu/~anderson},
  gr-qc/0001018, to appear in Annales Henri Poincar\'e, 2000.

\bibitem{AndChDiss}
L.~Andersson and P.T. Chru\'sciel, \emph{On asymptotic behaviour of solutions
  of the constraint equations in general relativity with ``hyperboloidal
  boundary conditions''}, Dissert. Math. \textbf{355} (1996), 1--100.

\bibitem{AChF}
L.~Andersson, P.T. Chru\'sciel, and H.~Friedrich, \emph{On the regularity of
  solutions to the {Y}amabe equation and the existence of smooth hyperboloidal
  initial data for {E}insteins field equations}, Comm. Math. Phys. \textbf{149}
  (1992), 587--612.

\bibitem{AndDahl}
L.~Andersson and M.~Dahl, \emph{Scalar curvature rigidity for asymptotically
  locally hyperbolic manifolds}, Annals of Global Anal.\ and Geom. \textbf{16}
  (1998), 1--27, dg-ga/9707017.

\bibitem{aronszajn}
N.~Aronszajn, \emph{A unique continuation theorem for solutions of elliptic
  partial differential equations or inequalities of second order}, Jour.\ de
  Math. \textbf{XXXVI} (1957), 235--249.

\bibitem{AshtekarDas}
A.~Ashtekar and S.~Das, \emph{Asymptotically anti-de {S}itter space-times:
  Conserved quantities}, Class.\ Quantum Grav. \textbf{17} (2000), L17--L30,
  hep-th/9911230.

\bibitem{AshtekarMagnonAdS}
A.~Ashtekar and A.~Magnon, \emph{Asymptotically anti--de {S}itter space-times},
  Classical Quantum Gravity \textbf{1} (1984), L39--L44.

\bibitem{bartnik:mass}
R.~Bartnik, \emph{The mass of an asymptotically flat manifold}, Comm. Pure
  Appl. Math. \textbf{39} (1986), 661--693.

\bibitem{Baum}
H.~Baum, \emph{Complete {R}iemannian manifolds with imaginary {K}illing
  spinors}, Ann.\ Global Anal.\ Geom. \textbf{7} (1989), 205--226.

\bibitem{BeigKomar}
R.~Beig, \emph{{Arnowitt--Deser--Misner energy and $g_{00}$}}, Phys.\ Lett.
  \textbf{69A} (1978), 153--155.

\bibitem{BeigSimon}
R.~Beig and W.~Simon, \emph{Proof of a multipole conjecture due to {Geroch}},
  Commun.\ Math.\ Phys. \textbf{78} (1980), 75--82.

\bibitem{BeigSimon3}
\bysame, \emph{On the uniqueness of static perfect--fluid solutions in general
  relativity}, Commun.\ Math.\ Phys. \textbf{144} (1992), 373--390.

\bibitem{BGH}
W.~Boucher, G.W. Gibbons, and G.T. Horowitz, \emph{Uniqueness theorem for
  anti--de {S}itter spacetime}, Phys.\ Rev.\ D \textbf{30} (1984), 2447--2451.

\bibitem{Bray:thesis}
H.~Bray, \emph{The {P}enrose inequality in general relativity and volume
  comparison theorem involving scalar curvature}, Ph.D. thesis, Stanford
  University, 1997.

\bibitem{Bray:preparation2}
\bysame, \emph{Proof of the {R}iemannian {P}enrose conjecture using the
  positive mass theorem}, math.DG/9911173, 1999.

\bibitem{BHKJ}
D.~Brill, G.T. Horowitz, D.~Kastor, and J.~Traschen, \emph{Testing cosmic
  censorship with black hole collisions}, Phys.\ Rev. \textbf{D49} (1994),
  840--852, gr-qc/9307014.

\bibitem{BLP}
D.~Brill, J.~Louko, and P.~Peldan, \emph{Thermodynamics of (3+1)-dimensional
  black holes with toroidal or higher genus horizons}, Phys.\ Rev. \textbf{D56}
  (1997), 3600--3610, gr-qc/9705012.

\bibitem{bunting:masood}
G.~Bunting and A.K.M. Masood{--ul--A}lam, \emph{Nonexistence of multiple black
  holes in asymptotically euclidean static vacuum space-time}, Gen.\ Rel.\
  Grav. \textbf{19} (1987), 147--154.

\bibitem{CarterJMP}
B.~Carter, \emph{Killing horizons and orthogonally transitive groups in
  space--time}, Jour.\ Math.\ Phys. \textbf{10} (1969), 70--81.

\bibitem{CarterHI}
\bysame, \emph{The general theory of the mechanical, electromagnetic and
  thermodynamic properties of black holes}, General relativity (S.W. Hawking
  and W.~Israel, eds.), Cambridge University Press, Cambridge, 1979,
  pp.~294--369.

\bibitem{ChAIHP}
P.T. Chru\'sciel, \emph{On the relation between the {Einstein} and the {Komar}
  expressions for the energy of the gravitational field}, Ann. Inst. H.
  Poincar\'e \textbf{42} (1985), 267--282.

\bibitem{ChErice}
\bysame, \emph{Boundary conditions at spatial infinity from a hamiltonian point
  of view}, Topological Properties and Global Structure of Space--Time (P.\
  Bergmann and V.\ de~Sabbata, eds.), Plenum Press, New York, 1986, {URL}
  \url{http://www.phys.univ-tours.fr/~piotr/scans}, pp.~49--59.

\bibitem{Chnohair}
\bysame, \emph{``{N}o {H}air'' {T}heorems -- folklore, conjectures, results},
  Differential Geometry and Mathematical Physics (J.~Beem and K.L. Duggal,
  eds.), Cont.\ Math., vol. 170, AMS, Providence, 1994, pp.~23--49,
  gr--qc/9402032.

\bibitem{Chstatic}
\bysame, \emph{The classification of static vacuum space--times containing an
  asymptotically flat spacelike hypersurface with compact interior}, Class.
  Quantum Grav. \textbf{16} (1999), 661--687, gr-qc/9809088.

\bibitem{Chstaticelvac}
\bysame, \emph{Towards the classification of static electro--vacuum
  space--times containing an asymptotically flat spacelike hypersurface with
  compact interior}, Class. Quantum Grav. \textbf{16} (1999), 689--704,
  gr-qc/9810022.

\bibitem{ChDGH}
P.T. Chru\'sciel, E.~Delay, G.~Galloway, and R.~Howard, \emph{Regularity of
  horizons and the area theorem}, Annales Henri Poincar\'e \textbf{in press}
  (2000), gr-qc/0001003.

\bibitem{CJK}
P.T. Chru\'sciel, J.~Jezierski, and J.~Kijowski, \emph{A {H}amiltonian
  framework for field theories in the radiating regime}, in preparation.

\bibitem{ForMac}
J.M. Foyster and C.B.G. McIntosh, \emph{A class of solutions of {E}instein's
  equations which admit a $3$-parameter group of isometries}, Commun.\ Math.\
  Phys. \textbf{27} (1972), 241--246.

\bibitem{Friedrich:adS}
H.~Friedrich, \emph{Einstein equations and conformal structure: Existence of
  anti-de-{Sitter}-type space-times}, Jour.\ Geom.\ and Phys. \textbf{17}
  (1995), 125--184.

\bibitem{GSWW2}
G.J. Galloway, K.~Schleich, D.M. Witt, and E.~Woolgar, \emph{The {AdS/CFT}
  correspondence conjecture and topological censorship},  (1999),
  hep-th/9912119.

\bibitem{GSWW}
\bysame, \emph{Topological censorship and higher genus black holes}, Phys. Rev.
  \textbf{D60} (1999), 104039, gr-qc/9902061.

\bibitem{Gannon2}
D.~Gannon, \emph{Singularities in nonsimply connected space-times}, Jour.\
  Math.\ Phys. \textbf{16} (1975), 2364--2367.

\bibitem{Gannon1}
\bysame, \emph{On the topology of spacelike hypersurfaces, singularities, and
  black holes}, Gen.\ Rel. Grav. \textbf{7} (1976), 219--232.

\bibitem{Geroch:limits}
R.~Geroch, \emph{Limits of spacetimes}, Commun.\ Math.\ Phys. \textbf{13}
  (1969), 180--193.

\bibitem{Geroch:extraction}
\bysame, \emph{Energy extraction}, Ann.\ New York Acad.\ Sci. \textbf{224}
  (1973), 108--117.

\bibitem{GibbonsGPI}
G.~Gibbons, \emph{Gravitational entropy and the inverse mean curvature flow},
  Class.\ Quantum Grav. \textbf{16} (1999), 1677--1687.

\bibitem{GHHP}
G.W. Gibbons, S.W. Hawking, G.T. Horowitz, and M.J. Perry, \emph{Positive mass
  theorem for black holes}, Commun. Math. Phys. \textbf{88} (1983), 295--308.

\bibitem{GT}
D.~Gilbarg and N.S. Trudinger, \emph{Elliptic partial differential equations of
  second order}, Springer, Berlin, 1983.

\bibitem{Goenner2}
H.~G\"onner and J.~Stachel, \emph{Einstein tensor and $3$-parameter groups of
  isometries with $2$-dimensional orbits}, Jour.\ Math.\ Phys. \textbf{11}
  (1970), 3358--3370.

\bibitem{SWH}
S.W. Hawking, \emph{Gravitational radiation in an expanding universe}, Jour.\
  Math.\ Phys. \textbf{9} (1968), 598--604.

\bibitem{HdS}
\bysame, \emph{The boundary conditions for gauged supergravity}, Phys. Lett. B
  \textbf{126} (1983), 175--177.

\bibitem{HT}
M.~Henneaux and C.~Teitelboim, \emph{Asymptotically anti--de {S}itter spaces},
  Commun.\ Math.\ Phys. \textbf{98} (1985), 391--424.

\bibitem{HorowitzMyers}
G.T. Horowitz and R.C. Myers, \emph{The {AdS/CFT} correspondence and a new
  positive energy conjecture for general relativity}, Phys. Rev. \textbf{D59}
  (1999), 026005 (12 pp.).

\bibitem{HI1}
G.~Huisken and T.~Ilmanen, \emph{The {Riemannian P}enrose inequality}, Int.\
  Math.\ Res.\ Not. \textbf{20} (1997), 1045--1058.

\bibitem{HI2}
\bysame, \emph{The inverse mean curvature flow and the {Riemannian P}enrose
  inequality}, Jour.\ Diff.\ Geom. (1999), in press.

\bibitem{Israel:vacuum}
W.~Israel, \emph{Event horizons in static vacuum space-times}, Phys. Rev.
  \textbf{164} (1967), 1776--1779.

\bibitem{JangWald}
P.S. Jang and R.M. Wald, \emph{The positive energy conjecture and the cosmic
  censor hypothesis}, Jour. Math.\ Phys \textbf{18} (1977), 41--44.

\bibitem{Kannar:adS}
J.~K{\'a}nn{\'a}r, \emph{Hyperboloidal initial data for the vacuum {E}instein
  equations with cosmological constant}, Class.\ Quantum Grav. \textbf{13}
  (1996), 3075--3084.

\bibitem{KastorTraschen}
D.~Kastor and J.~Traschen, \emph{Cosmological multi--black hole solutions},
  Phys.\ Rev. \textbf{D47} (1993), no.~10, 5370--5375, hep-th/9212035.

\bibitem{KennefickMurchadha}
D.~Kennefick and N.~{\'O} Murchadha, \emph{Weakly decaying asymptotically flat
  static and stationary solutions to the {E}instein equations}, Class.\ Quantum
  Grav. \textbf{12} (1995), 149--158.

\bibitem{Kijowskiold}
J.~Kijowski, \emph{On a new variational principle in general relativity and the
  energy of the gravitational field}, Gen. Rel. Grav. \textbf{9} (1978),
  857--877.

\bibitem{KijowskiGRG}
\bysame, \emph{A simple derivation of canonical structure and quasi-local
  {Hamiltonians} in general relativity}, Gen. Rel. Grav. \textbf{29} (1997),
  307--343.

\bibitem{KijowskiTulczyjew}
J.~Kijowski and W.M. Tulczyjew, \emph{A symplectic framework for field
  theories}, Lecture Notes in Physics, vol. 107, Springer, New York,
  Heidelberg, Berlin, 1979.

\bibitem{Kottler}
F.~Kottler, \emph{{\"Uber die physikalischen Grundlagen der Einsteinschen
  Gravitationstheorie}}, Annalen der Physik \textbf{56} (1918), 401--462.

\bibitem{LL}
L.~Lindblom, \emph{Static uniform-density stars must be spherical in general
  relativity}, Jour.\ Math.\ Phys. \textbf{29} (1988), 436--439.

\bibitem{Magnon}
A.~Magnon, \emph{On {K}omar integrals in asymptotically anti--de {S}itter
  space-times}, Jour.\ Math.\ Phys. \textbf{26} (1985), 3112--3117.

\bibitem{Manncqg}
R.~B. Mann, \emph{Pair production of topological anti--de {S}itter black
  holes}, Class.\ Quantum Grav. \textbf{14} (1997), L109--L114, gr-qc/9607071.

\bibitem{Nariai}
H.~Nariai, \emph{On a new cosmological solution of {E}instein's field equations
  of gravitation}, Sci. Rep. Tohoku Univ. Ser. I. \textbf{35} (1951), 62--67.

\bibitem{penrose:scri}
R.~Penrose, \emph{Zero rest-mass fields including gravitation}, Proc.\ Roy.\
  Soc.\ London \textbf{A284} (1965), 159--203.

\bibitem{PenroseSCC}
\bysame, \emph{Gravitational collapse --- the role of general relativity}, Riv.
  del Nuovo Cim. \textbf{1 (numero speziale)} (1969), 252--276.

\bibitem{SimonBeig}
W.~Simon and R.~Beig, \emph{The multipole structure of stationary
  space--times}, Jour.\ Math.\ Phys. \textbf{24} (1983), 1163--1171.

\bibitem{deSitter1917b}
W.~De Sitter, \emph{On the curvature of space}, Proc.\ Kon.\ Ned.\ Akad.\ Wet.
  \textbf{20} (1917), 229--243.

\bibitem{Stanciulescu}
C.~Stanciulescu, \emph{Spherically symmetric solutions of the vacuum {E}instein
  field equations with positive cosmological constant}, 1998, Diploma Thesis,
  University of Vienna.

\bibitem{Vanzo}
L.~Vanzo, \emph{Black holes with unusual topology}, Phys. Rev. \textbf{D56}
  (1997), 6475--6483, gr-qc/9705004.

\bibitem{Vishveshwara}
C.V. Vishveshwara, \emph{Generalization of the ``{S}chwarzschild surface'' to
  arbitrary static and stationary metrics}, Jour.\ Math.\ Phys. \textbf{9}
  (1968), 1319--1322.

\bibitem{Walker}
M.~Walker, \emph{Bloc diagrams and the extension of timelike two--surfaces},
  Jour.\ Math.\ Phys. \textbf{11} (1970), 2280--2286.

\bibitem{WittenYau}
E.~Witten and S.T. Yau, \emph{Connectedness of the boundary in the {AdS/CFT}
  correspondence},  (1999), hep-th/9910245.

\bibitem{Woolgar:area}
E.~Woolgar, \emph{Bounded area theorems for higher genus black holes}, Class.\
  Quantum Grav. \textbf{16} (1999), 3005--3012, gr-qc/9906096.

\end{thebibliography}
\end{document}